\documentstyle[aps,preprint,prl,epsfig]{revtex}
\begin{document}

\title{Anharmonic double-phonon excitations 
in the interacting boson model}
\author{
J.E.~Garc\'{\i}a--Ramos$^{1,2,}$\footnote{Postdoctoral research fellow
of the Fund of Scientific Research, Flanders, Belgium},
J.M.~Arias$^2$,
and
P.~Van Isacker$^3$}
\address{$^1$Institute for Theoretical Physics,
Vakgroep Subatomaire en Stralingsfysica,
Proeftuinstraat 86, B-9000 Gent, Belgium}

\address{$^2$Departamento de F\'{\i}sica At\'omica, Molecular y Nuclear,
Universidad de Sevilla, Aptdo. 1065, 41080 Sevilla, Spain}

\address{$^3$Grand Acc\'el\'erateur National d'Ions Lourds,
B.P.~5027, F-14076 Caen Cedex 5, France}
\date{\today}
\maketitle

\begin{abstract}
Double-$\gamma$ vibrations in deformed nuclei
are analyzed in the context of the interacting boson model.
A simple extension of the original version of the model
towards higher-order interactions
is required to explain the observed anharmonicities
of nuclear vibrations.
The influence of three- and four-body interactions
on the moments of inertia of ground- and $\gamma$-bands,
and on the relative position of single-$\gamma$ and double-$\gamma$ bands
is studied in detail.
As an example of a realistic calculation,
spectra and transitions of the highly $\gamma$-anharmonic nuclei
$^{164}$Dy, $^{166}$Er, and $^{168}$Er
are interpreted in this approach.
\end{abstract}

%\pacs{PACS numbers: 21.60 -n, 21.60 Fw, 21.60 Ev}
\vspace{1cm}
~~~~~~{\bf PACS numbers: 21.60 -n, 21.60 Fw, 21.60 Ev}

\draft
\section{Introduction}
\label{intro}
Vibrational degrees of freedom in atomic nuclei
can be described in terms of phonon excitations
that arise from nuclear shape oscillations.
Vibrations of nuclei with ellipsoidal symmetry
can be of two types~\cite{Bohr75}:
$\beta$ vibrations which preserve axial symmetry
and give rise to a band with $K=0$,
and $\gamma$ vibrations which break axial symmetry
and yield a $K=2$ band,
where $K$ is the projection of the angular momentum
on the axis of symmetry.
At the experimental level,
$\gamma$ bands have been identified in many well-deformed nuclei;
in contrast, the identification of $\beta$ bands
is still full of questions and difficulties.
This is mainly because,
when the energy surface has a well-deformed minimum in $\beta$
but is rather flat in $\gamma$,
the $\beta$-band increases in excitation energy
and approaches the energy region
where other degrees of freedom are important. 
In that case band mixing may occur and can give rise to non-pure 
structures with decay patterns difficult to identify as those of a
$\beta$ band \cite{Garr97b}.

Since single-$\gamma$ excitations are very well established,
it is natural to look for double-$\gamma$ vibrations
and to develop models that can deal
with such multi-phonon excitations.
Double-$\gamma$ excitations correspond to
$K^\pi=0^+$ and $K^\pi=4^+$ bands
which are the anti-parallel and parallel combinations
of single-$\gamma$ phonons, respectively.
The experimental identification of two-$\gamma$ states
in deformed nuclei is difficult
because their expected excitation energy
is around the pairing gap
and hence they can mix strongly with two-quasiparticle excitations.
However, recent experimental improvements
in nuclear spectroscopy
following Coulomb excitation~\cite{Fahl88},
inelastic neutron scattering~\cite{Belg96},
and thermal-neutron capture~\cite{Born93}
have made possible the study of highly excited low-spin states.
Many states have been proposed
as possible candidates of double-$\gamma$ vibrations.
There is, however, some controversy about their interpretation.
The author of reference~\cite{Burk94}
claims that some of the presumed double-$\gamma$ states
can be interpreted as single hexadecapole-phonon excitations.
In fact, to identify the band-head of a double-$\gamma$ band
it is not sufficient to analyze just $B(E2)$ values;
data from single-nucleon transfer reactions,
$\beta$-decay studies,
and inelastic scattering experiments
must be considered as well.
One of the key properties
to disregard a band as a double-$\gamma$ band
is the fact that its members, in first order,
cannot be populated in single-nucleon transfer reactions.
Many examples of $K^\pi=4^+$ states
that are identified as double-$\gamma$ excitations
but are strongly populated in single-nucleon transfer reactions,
can be found in the literature:
$^{158}$Gd,
$^{162}$Dy,
$^{172}$Yb,
$^{176,178}$Hf,
and $^{190,192}$Os.
Since the double-phonon character
of the states in question is in doubt,
they are not considered here.
However, some candidates
seem to have a genuine double-phonon nature.
Such is the case with $^{164}$Dy and $^{166-68}$Er.
In particular, in reference~\cite{Corm97}
a $K^\pi=4^+$ state in $^{164}$Dy at $2.173$ MeV
is found to exhibit all properties of a
double-$\gamma$ band.
In references~\cite{Fahl96,Garr97}
the observation is reported of $K^\pi=0^+$ and $K^\pi=4^+$
double-$\gamma$ states in $^{166}$Er,
at energies of $1.949$ MeV and $2.029$ MeV, respectively.
Finally, in $^{168}$Er a $K^\pi=4^+$ double-$\gamma$ excitation
is identified at an energy of $2.055$ MeV~\cite{Born91}.

One of the most striking features
of the observed double-$\gamma$ bands
is their high anharmonicity,
{\it i.e.}~the ratio of double-$\gamma$ over single-$\gamma$ energy
is different from $2$
and ranges from $2.5$ to $2.8$.
This information is very important
since it provides a stringent test of nuclear models.
The nuclei $^{164}$Dy and $^{166,168}$Er
have been interpreted in the context of many different models
such as the quasi-phonon model~\cite{Solo94,Solo95},
the geometrical model~\cite{Bohr82},
the multi-phonon model~\cite{Jamm88},
the self-consistent collective-coordinate method~\cite{Mats86,Mats87},
and the $sdg$-IBM~\cite{Yosh86,Yosh88},
and it is now of interest to revisit these models
in connection with anharmonic vibrational behavior.

In this paper the simplest version
of the interacting boson model (IBM)~\cite{Iach87}
is extended by adding to the usual Hamiltonian
higher-order interactions between the bosons 
with the purpose of creating a framework
that accommodates the high anharmonicities
observed in $^{164}$Dy and $^{166,168}$Er.
The structure of the paper is as follows.
First, the IBM is reviewed
with special reference to its harmonic character.
In section~\ref{3b} the inclusion of three-body terms
in the Hamiltonian is discussed.
The introduction of four-body terms
is presented in section~\ref{su3-4b}
and some analytic results are pointed out.
In section~\ref{4b} a detailed study of possible four-body terms
is carried out
and realistic calculations
for $^{164}$Dy and $^{166,168}$Er
are presented.
Finally, in section~\ref{conclu}
the conclusions of this work are made.

\section{The IBM-1 as a harmonic model}
\label{ibm1}
The IBM describes low-lying collective excitations in even--even nuclei
in terms of monopole ($s$) and quadrupole ($d$) bosons~\cite{Iach87}. 
The boson number that corresponds to a given nucleus
equals half the number of valence nucleons ($N=n/2$).
The rotationally invariant and number-conserving boson Hamiltonian
usually includes up to two-body interactions between the bosons
although higher-order terms can be added in principle.
The most general two-body IBM Hamiltonian can be written
in a multipole expansion as
\begin{equation}
\label{Ham1}
\hat H=
\varepsilon_s \hat n_s+
\varepsilon_d \hat n_d+
\kappa_0 \hat P^\dagger\,\hat P+
\kappa_1 \hat L\cdot\hat L+
\kappa_2 \hat Q\cdot\hat Q+
\kappa_3 \hat T_3\cdot\hat T _3+
\kappa_4 \hat T_4\cdot\hat T _4,
\end{equation}
where $\hat n_s$ and $\hat n_d$ 
are the $s$- and $d$-boson number operators, respectively,
and
\begin{eqnarray}
\hat{P}^\dagger&=&
{1\over2} d^\dag\cdot d^\dag-  
{1\over2} s^\dag\cdot s^\dag,
\label{P}\\
\hat L&=&
\sqrt{10}(d^\dag\times\tilde{d})^{(1)},
\label{L}\\
\hat Q&=&
s^\dag\tilde d+
d^\dag\tilde s+\chi(d^\dag\times\tilde d)^{(2)},
\label{Q}\\
\hat T_3&=&
(d^\dag\times\tilde{d})^{(3)},
\label{t3}\\
\hat T_4&=&
(d^\dag\times\tilde{d})^{(3)}.
\label{t4}
\end{eqnarray} 
The symbol $\cdot$ represents the scalar product;
in this paper the scalar product of two operators
with angular momentum $L$ is defined as 
$\hat T_L\cdot\hat T_L=\sum_M (-1)^M\hat T_{LM}\hat T_{L-M}$
where $\hat T_{LM}$ corresponds to the $M$ component
of the operator $\hat T_{L}$.
In the previous equations
the operator 
$\tilde\gamma_{\ell m}=(-1)^m\gamma_{\ell -m}$
(where $\gamma$ refers to $s$ or $d$)
is introduced so that the annihilation operator
verifies the appropriate properties under spatial rotations.

It is not {\it a priori} clear
to what extend $\beta$ and $\gamma$ vibrations
are anharmonic in the IBM
even if one just considers the
Hamiltonian~(\ref{Ham1}) with up to two-body interactions.
A partial analysis of this problem
was given in reference~\cite{Garc98c}.
There, the authors find that  
the IBM in its simplest version is a harmonic model
in the limit of infinite boson number $N$
and even for finite $N$
the model cannot accommodate large anharmonicity
if one considers up to two-body interactions;
only the interplay between one+two-body terms
and higher-order interactions
can induce, in principle, a sizeable anharmonicity
in the double-phonon excitations.  
The reason why one-body terms and two-body interactions
cannot create a large anharmonicity
can be understood as follows.
If one considers a Hamiltonian
with one parameter that controls
the ratio of the strength of the one-body energies
and the two-body interactions,
\begin{equation}
\label{ham-phase}
\hat H=
(1-\xi)\Big(
\varepsilon_s \hat n_s+ 
\varepsilon_d \hat n_d\Big)+  
\xi\Big(
\kappa_0 \hat P^\dag\,\hat P+
\kappa_1 \hat L\cdot\hat L+
\kappa_2 \hat Q\cdot\hat Q+
\kappa_3 \hat T_3\cdot\hat T_3+
\kappa_4 \hat T_4\cdot\hat T_4\Big),
\end{equation}
where $\xi$ ranges from $0$ to $1$, 
one finds two `phases' separated by a critical value,
$\xi_{\rm c}$:
a first phase
where the one-body term
plays the main role ($\xi<\xi_{\rm c}$)
and a second phase
where the two-body interaction
is the driving force ($\xi>\xi_{\rm c}$).
The crucial point is that
the separation between the two phases is very sharp~\cite{Iach98}
and essentially no interplay
between one- and two-body terms can be found.
Since, to a good approximation,
the force is either one body or two body but not both,
harmonic behavior cannot be avoided.  

The inclusion of high-order interactions
in a system with a high boson number
also leads to a harmonic description.
Only for finite boson number
the interplay between one+two-body terms
and higher-order interactions
can induce an anharmonicity
in the double-phonon excitations
that is comparable to the observed one.
These ideas will be used as a guideline in the following sections.

To carry out a quantitative study of anharmonicities,
it is convenient to define a ratio
between single- and double-phonon excitation energies.
Because the experimental situation for $\beta$ excitations
is not clear
we concentrate on $\gamma$ vibrations
and define energy ratios for $\gamma$ phonons only
(although similar definitions can be given for $\beta$ phonons):
\begin{equation}
R_0^\gamma\equiv
{{E_{\rm x}(0^+_{\gamma\gamma})}
\over
{E_{\rm x}(2^+_\gamma)-E_{\rm x}(2^+_1)}},
\qquad
R_4^\gamma\equiv
{{E_{\rm x}(4^+_{\gamma\gamma})-E_{\rm x}(4^+_1)}
\over
{E_{\rm x}(2^+_\gamma)-E_{\rm x}(2^+_1)}},
\label{ratios}
\end{equation}
where $0^+_{\gamma\gamma}$ and $4^+_{\gamma\gamma}$
are the band heads of the $K^\pi=0^+$
and $K^\pi=4^+$ double-$\gamma$ bands,
respectively,
and $E_{\rm x}$ stands for excitation energy. This particular
definition removes any rotational influence.

\section{Three-body Hamiltonians}
\label{3b}
Let us consider in the following a Hamiltonian that 
includes a quadrupole term,
a rotational $\hat L^2$ term,
and three-body interactions between the $d$ bosons,
\begin{equation}
\hat H=
-\kappa\hat Q\cdot\hat Q+ 
\kappa'\hat L\cdot\hat L+
\sum_{k l}\theta_l
\left((d^\dag\times d^\dag)^{(k)}\times d^\dag\right)^{(l)}
\cdot
\left((\tilde d\times\tilde d)^{(k)}\times\tilde d\right)^{(l)},
\label{ham}
\end{equation}
where $-\kappa=\kappa_2$, $\kappa'=\kappa_1$.
Five independent three-body $d$-boson interactions exist
which have $l=0$, 2, 3, 4, and 6.
Interactions with the same $l$ but different $k$
are not independent
but differ by a normalization factor only~\cite{Isac81}.
The combinations
$(k,l)=(2,0)$, (0,2), (2,3), (2,4), and (4,6) are chosen here.

The Hamiltonian~(\ref{ham})
is certainly not the most general one+two+three-body Hamiltonian
that can be considered.
Notably, a vibrational term $\epsilon_d\hat n_d$
which dominates in spherical nuclei is omitted
since it is thought of lesser importance
in the deformed nuclei considered here.
And, of all possible three-body interactions (seventeen terms),
only those between the $d$ bosons are retained here
since they are the more efficient terms to produce anharmonicity 
\cite{unpu}.

\begin{figure}[hbt]
\begin{center}
\mbox{\epsfig{file=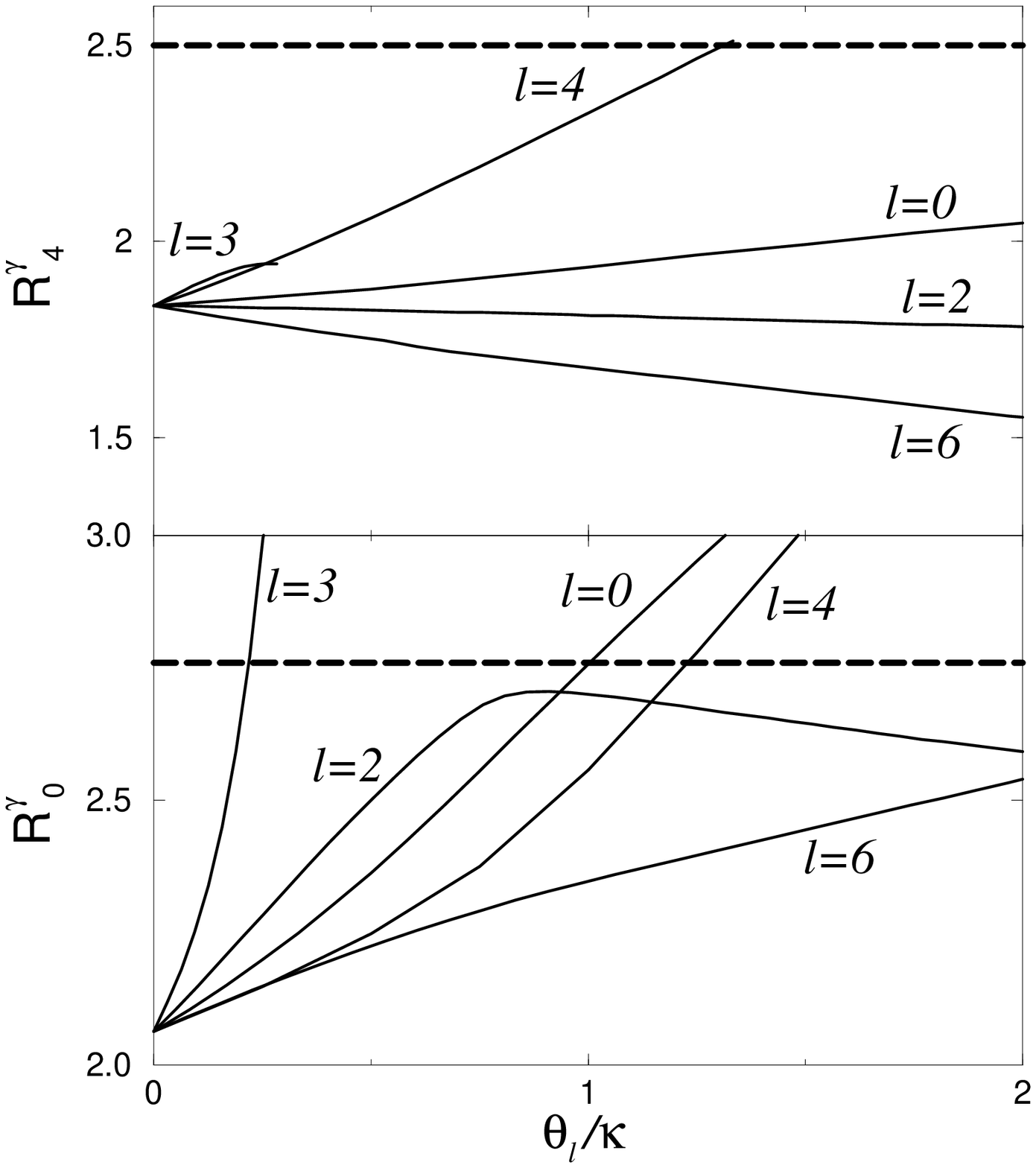,height=15.0cm,angle=-90}}
\end{center}
\caption{The ratios $R_K^\gamma$ (as defined in the text)
as a function of $\theta_l/\kappa$ for different $l$.
The Hamiltonian~(\ref{ham}) is used with $\chi=-0.5$;
the boson number is $N=15$.
The dashed lines give the experimental values
for the corresponding ratios in $^{166}$Er.}
\label{fig-3b-ratios}
\end{figure}

\begin{figure}[hbt]
\begin{center}
\mbox{\epsfig{file=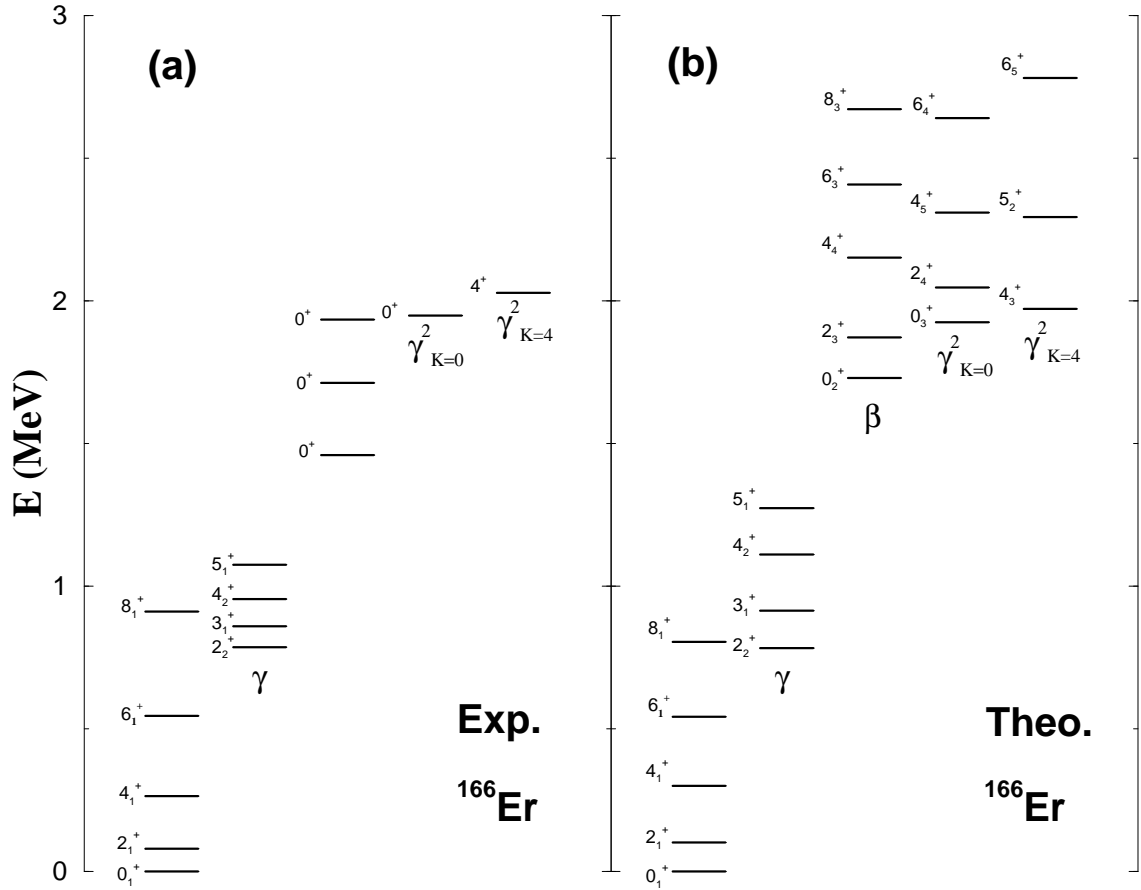,height=15.0cm,angle=-90}}
\end{center}
\caption{Experimental (a) and calculated (b) spectrum for $^{166}$Er. 
The theoretical results are obtained
with the Hamiltonian~(\ref{ham})
with $\kappa=23.8$ keV, $\chi=-0.55$, $\kappa'=-1.9$ keV,
and $\theta_4=31.3$ keV. The boson number is $N=15$.}
\label{fig-spec-3b}
\end{figure}

For the discussion of the anharmonicities of $\gamma$ vibrations
we study the behavior of the energy ratios~(\ref{ratios}) as a
function of the ratio $\theta_l/\kappa$.
The identification of the states
$0^+_{\gamma\gamma}$ and $4^+_{\gamma\gamma}$
is based on the $B(E2)$ values
for decay into the single-$\gamma$ states.
In figure~\ref{fig-3b-ratios}\footnote{Note that this figure differs
from figure 2 in reference \cite{Garc00b} in some scale factors due to
an error in the definition of $\theta_l$.}
the influence of the various three-body interactions
is shown for a typical value of $\chi$ ($\chi=-0.5$)
and for $N=15$ bosons.
It is seen that a $\gamma$-vibrational anharmonic behavior
is obtained
which can be different for the $K^\pi=0^+$ and $K^\pi=4^+$ bands
({\it e.g.} positive for the former while negative for the latter.)
Care has been taken to plot results only up to values of $\theta_l$
that do not drastically alter the character of rotational spectrum;
beyond these values, the three-body interaction,
being of highest order in the Hamiltonian~(\ref{ham}),
becomes dominant.
Also shown in figure~\ref{fig-3b-ratios}
are the ratios $R_K^\gamma$
as observed in $^{166}$Er~\cite{Fahl96,Garr97},
$R_0^\gamma=2.76$ and $R_4^\gamma=2.50$.
Figure~\ref{fig-spec-3b}
shows the experimental spectrum of $^{166}$Er~\cite{Fahl96,Garr97}
and compares it to the eigenspectrum of Hamiltonian~(\ref{ham})
with an $l=4$ three-body interaction.
The parameters are $\kappa=23.8$ keV, $\chi=-0.55$,
$\kappa'=-1.9$ keV, and $\theta_4=31.3$ keV\footnote{
Note that the value $\theta_4$ is $1/3$ of that
given in reference~\cite{Garc00b}
which has an error in its definition.
The results shown in that paper are correct
after a simple rescaling of the parameter.},
with boson number $N=15$.
With these values the
calculated excitation energies of the double-$\gamma$ band heads
are 1926 keV and 1972 keV for the $K^\pi=0^+$ and $K^\pi=4^+$ levels,
respectively, leading to the ratios
$R_0^\gamma=2.82$ and $R_4^\gamma=2.45$,
in excellent agreement with observation.
Note, however, that although all $\gamma$-band heads
are well reproduced by the calculation,
problems arise for the moments of inertia,
in particular of the $\gamma$ band. In next sections we will come back
on the moments of inertia to see that three-body Hamiltonians
provide a very poor description of them.

\section{$SU(3)$ Hamiltonians with up to four-body interactions}
\label{su3-4b}
In the previous section
and in reference~\cite{Garc00b}
a very good description of double-phonon excitation energy
has been obtained, 
but at expense of spoiling the moments of inertia
of ground and $\gamma$ bands.
These drawbacks seem to be a general feature
of three-body Hamiltonians.
In this and the following section
it is shown that the drawbacks of three-body interactions
can be overcome by going to the next order.

Since $\gamma$ anharmonicity has been observed
exclusively in well-deformed nuclei,
it is appropriate to consider the problem
in the SU(3) limit of the IBM
which is suited to deal with nuclei
in this mass region~\cite{Cast88}.
Therefore, in a first approach, the $SU(3)$ limit is used
and later, in the next section,
these results are used as a guidance
in more realistic calculations. 

Let us consider the following Hamiltonian:
\begin{eqnarray}
\label{cas-ham}
\hat H&=&
a~\hat C_2[SU(3)]+
b_1~\hat C_3[SU(3)]+
b_2~\hat N \hat C_2[SU(3)]
\nonumber\\
&+&
c_1~\hat C_2[SU(3)]^2+
c_2~\hat N \hat C_3[SU(3)]+
c_3~\hat N^2 \hat C_2[SU(3)],
\end{eqnarray}
where $\hat C_n[SU(3)]$ stands for
the Casimir operator of order $n$ of $SU(3)$. Note the inclusion of
cubic terms for completeness in the $SU(3)$ analysis.
Since only the spectrum of a single nucleus
is of interest here,
the number of bosons can be fixed in every case
and the Hamiltonian~(\ref{cas-ham})
can be simplified by combining terms into a single one,
leaving a Hamiltonian with three
coefficients $a$, $b$, and $c$,
\begin{eqnarray}
\label{cas-ham2}
\hat H&=&
a~\hat C_2[SU(3)]+
b~\hat C_3[SU(3)]+
c~\hat C_2[SU(3)]^2,
\end{eqnarray}
with eigenvalues
\begin{eqnarray}
\label{eigen-cas}
\langle(\lambda,\mu)|\hat H|(\lambda,\mu)\rangle&=& 
a(\lambda^2+\mu^2+\lambda\mu+3\lambda+3\mu)\nonumber\\
&+&b(\lambda-\mu)(2\lambda+\mu+3)(\lambda+2\mu+3)\nonumber\\
&+&c(\lambda^2+\mu^2+\lambda\mu+3\lambda+3\mu)^2.
\end{eqnarray}
No quartic Casimir operator exists for $SU(3)$
because the number of independent Casimir operators
equals the number of labels
that characterize an irreducible representation.
The Hamiltonian~(\ref{cas-ham2})
has no rotational term $\hat L^2$
since of primary interest, at this point,
is the description of band-head energies
of single- and double-$\gamma$ excitations.

The definition of the energy ratios~(\ref{ratios})
must now be adapted to incorporate
the symmetry labeling of the states.
In the $SU(3)$ limit
the $\gamma$ band belongs to the $SU(3)$ representation $(2N-4,2)$
where $N$ is the number of bosons.
The double-$\gamma$ band with $K=4$
is contained in the $(2N-8,4)$ representation;
the double-$\gamma$ band with $K=0$
is predominantly contained in the $(2N-6,0)$ representation,
although an important component is in $(2N-8,4)$ \cite{Garc98c}.
In this section the energy ratios~(\ref{ratios})
are thus defined as follows:
\begin{equation}
R_0^{SU(3)\gamma}\equiv
{{E_{\rm x}(2N-6,0)}
\over
{E_{\rm x}(2N-4,2)}},
\qquad
R_4^{SU(3)\gamma}\equiv
{{E_{\rm x}(2N-8,4)}
\over
E_{{\rm x}}(2N-4,2)}.
\label{ratios-su3}
\end{equation}
Because no rotational term is included, 
the ratios~(\ref{ratios-su3}) can be compared directly
with equations~(\ref{ratios}).
In the following the effect of the different terms
in equation~(\ref{cas-ham2}) on the degree of anharmonicity
of the two-phonon excitation energy is analyzed.

\begin{figure}[hbt]
\begin{center}
\mbox{\epsfig{file=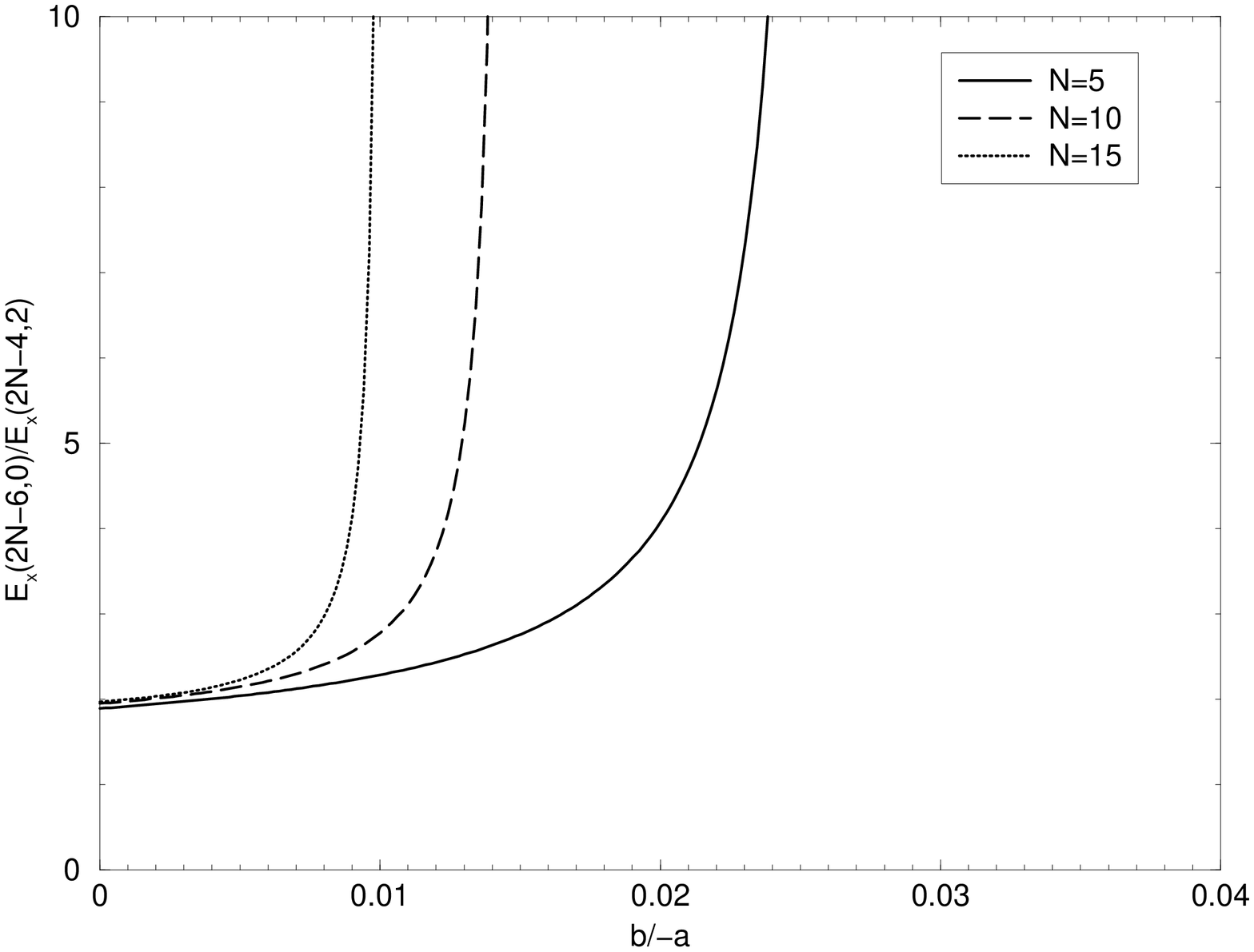,height=15.0cm,angle=-90}}
\end{center}
\caption{The ratio $R_0^{SU(3)\gamma}$ (as defined in the text)
for the Hamiltonian\\
$\hat H=a~\hat C_2[SU(3)]+b~\hat C_3[SU(3)]$
with $N=5$, $10$, and $15$.} 
\label{fig-ratio-3b}
\end{figure}

\begin{figure}[hbt]
\begin{center}
\mbox{\epsfig{file=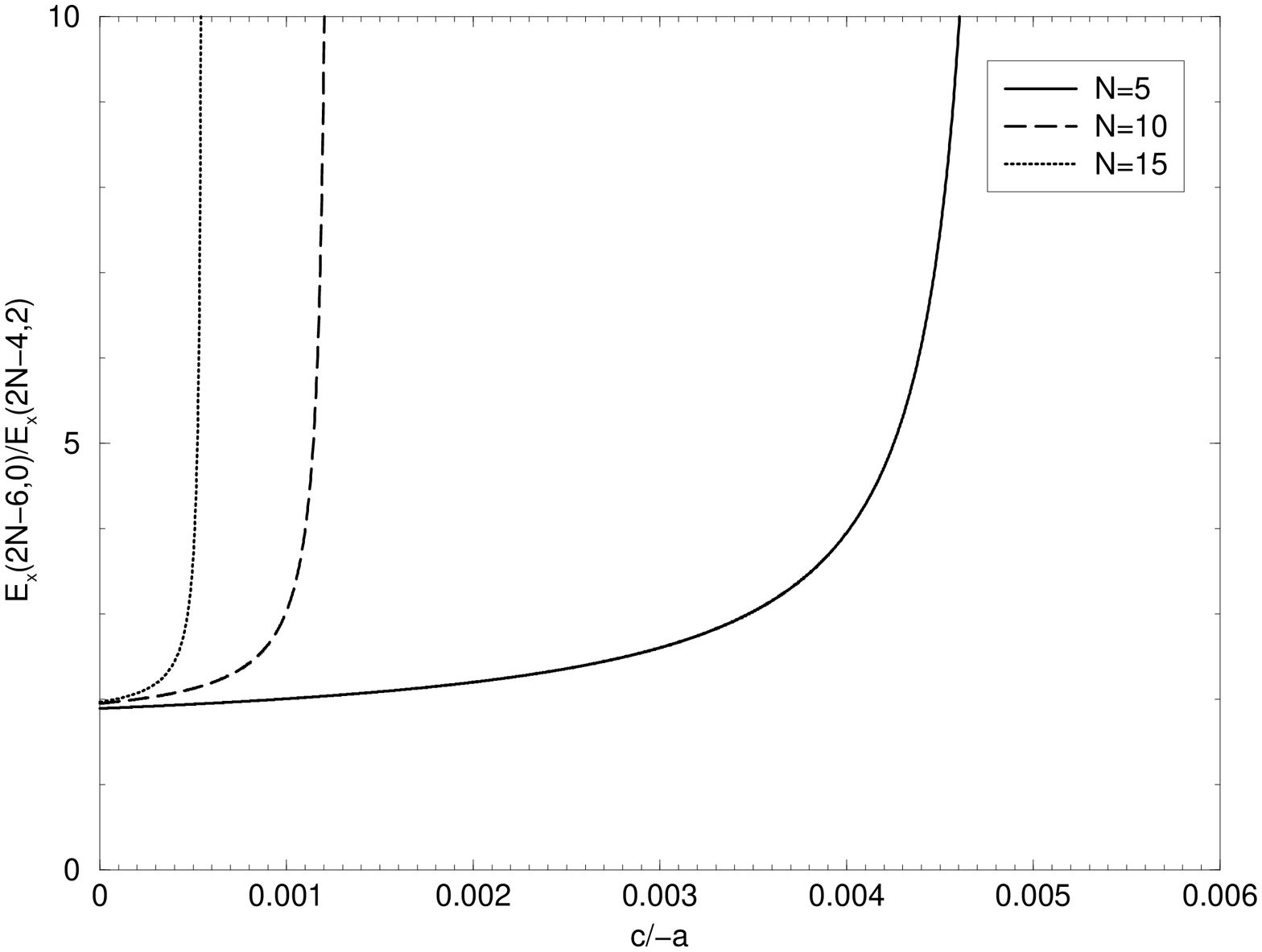,height=15.0cm,angle=-90}}
\end{center}
\caption{The ratio $R_0^{SU(3)\gamma}$ (as defined in the text)
for the Hamiltonian\\
$\hat H=a~\hat C_3[SU(3)]+c~\hat C_2[SU(3)]^2$
with $N=5$, $10$, and $15$.} 
\label{fig-ratio-60}
\end{figure}

\begin{figure}[hbt]
\begin{center}
\mbox{\epsfig{file=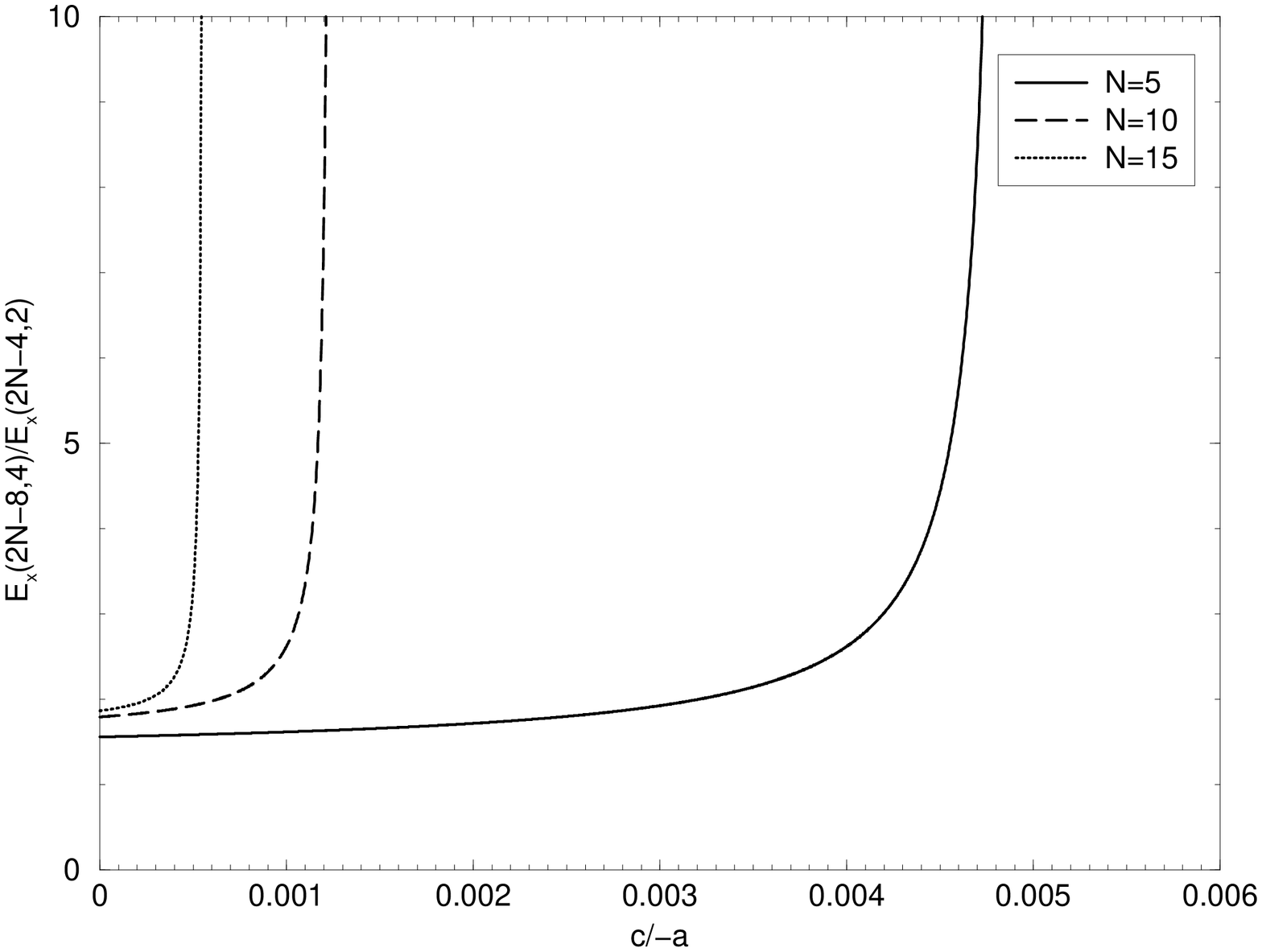,height=15.0cm,angle=-90}}
\end{center}
\caption{The ratio $R_4^{SU(3)\gamma}$ (as defined in the text)
for the Hamiltonian\\
$\hat H=a~\hat C_2[SU(3)]+c~\hat C_2[SU(3)]^2$
with $N=5$, $10$, and $15$.} 
\label{fig-ratio-84}
\end{figure}

\begin{itemize}
\item $a<0$, $b=0$, $c=0$.

This corresponds to the simplest version of IBM;
the energy ratios become
\begin{eqnarray}
\label{ratios-a}
R_0^{SU(3)\gamma}={{24N-18}\over{12N-6}},
\qquad
R_4^{SU(3)\gamma}={{24N-36}\over{12N-6}}.
\end{eqnarray}
An almost pure harmonic $\gamma$-vibrational spectrum is found
since the energy ratios~(\ref{ratios-a})
are only slightly lower than $2$.
This will be referred to as negative anharmonicity
as opposed to the positive anharmonicity
for energy ratios above $2$.

\item $a<0$, $b\neq 0$, $c=0$.

In this case the Hamiltonian~(\ref{cas-ham2})
is a combination of two- and three-body terms.
For given values of $a$ and $b$
and for a high enough boson number,
only the three-body part of the Hamiltonian is dominant
and a harmonic spectrum is recovered.
For obtaining an anharmonic spectrum
the values of the Hamiltonian parameters
are very constrained once the number of bosons has been fixed,
as the following analysis shows. 

Without lost of generality $a$ can be fixed to $a=-1$. 
The energy ratios are then
\begin{eqnarray}
\label{ratios-b1}
R_0^{SU(3)\gamma}&=&{{3-4N+b(24N^2-36N+27)}\over{(2N-1)(6bN+9b-1)}},\\
\label{ratios-b2}
R_4^{SU(3)\gamma}&=&2{{2N-3}\over{2N-1}}.
\end{eqnarray}
To keep the energy of the $\gamma$ excitation positive,
the value of $b$ has an upper limit
\begin{equation}
b<{1\over{6N+9}}.
\end{equation}
The value $b=1/(6N+9)$
leads to a divergence in $R_0^{SU(3)\gamma}$
and around this value anharmonic behavior is found. 
From equation~(\ref{ratios-b2}) one observes
that the behavior of the $(2N-8,4)$ representation
is completely harmonic and does not depend on $b$.
On the other hand,
from equation~(\ref{ratios-b1}) one sees
that a wide range of anharmonic ratios
is found for the $(2N-6,0)$ representation.
As an illustration, in figure~\ref{fig-ratio-3b}
equation~(\ref{ratios-b1}) is represented as a function of $b$,
for three values of $N$ ($5$, $10$, and $15$).
Only positive values of $b$ are plotted
because for the negative ones 
$R_0^{SU(3)\gamma}$ decreases smoothly
to the asymptotic values $1.27$, $1.58$, and $1.70$
for $N=5$, $10$, and $15$, respectively.    

The conclusion is that a Hamiltonian with $c=0$
does not agree with the experimental situation
observed in the mass region of well-deformed nuclei,
where the $\gamma^2_{K=4}$ state is highly anharmonic. 

\item $a<0$, $b=0$, $c\neq0$.

In this case the Hamiltonian~(\ref{cas-ham2})
is a combination of two- and four-body terms.
As in the previous case,
anharmonicity requires very constrained Hamiltonian parameters
once the number of bosons is fixed. 

Again, without loss of generality, we fix $a=-1$.
The energy ratios then read as follows:
\begin{eqnarray}
\label{ratios-c1}
R_0^{SU(3)\gamma}&=&{{(4N-3)(2c(4N^2-6N+9)-1)}\over{(2N-1)(8cN^2+6c-1)}},\\
\label{ratios-c2}
R_4^{SU(3)\gamma}&=&{{2(2N-3)(4c(2N^2-3N+9)-1)}\over{(2N-1)(8cN^2+6c-1)}}.
\end{eqnarray}
To keep the energy of the $\gamma$ excitation positive,
the value of $c$ has the upper limit
\begin{equation}
c<{1\over{8N^2+6}}.
\end{equation}
The value $c=1/(8N^2+6)$ produces a divergence
in the two energy ratios
and in its neighborhood highly anharmonic behavior is found.
In figures~\ref{fig-ratio-60} and \ref{fig-ratio-84}
are plotted the ratios
$R_0^{SU(3)\gamma}$ and $R_4^{SU(3)\gamma}$, respectively.
Again, for negatives values of $c$,
$R_0^{SU(3)\gamma}$ goes asymptotically
to the values $1.45$, $1.69$, and $1.78$
for $N=5$, $10$, and $15$, respectively,
while $R_4^{SU(3)\gamma}$
goes to $1.33$, $1.59$, and $1.71$.
These negative anharmonicities
have no phenomenological interest.

In this case the experimental situation
can be nicely described.
Both $\gamma^2_{K=0}$ and $\gamma^2_{K=4}$ states
can be accommodated in a anharmonic description.
The conclusion is thus that
a Hamiltonian with $\hat C[SU(3)]$ and $\hat C[SU(3)]^2$ terms
seems to be a good starting point
to treat the $\gamma$ anharmonicity in deformed nuclei.
\end{itemize}

The description presented here only provides band heads
and, to recover a rotational structure,
a $\hat L^2$ term must be included.
The rotational structure is the same in every band
because in the $SU(3)$ limit no mixing exists
between rotational and vibrational degrees of freedom.
In next subsection
these results are illustrated with a schematic calculation
for energies and transition probabilities.

\subsection{A schematic application}
\label{su3-app-sch}
\begin{figure}[hbt]
\begin{center}
\mbox{\epsfig{file=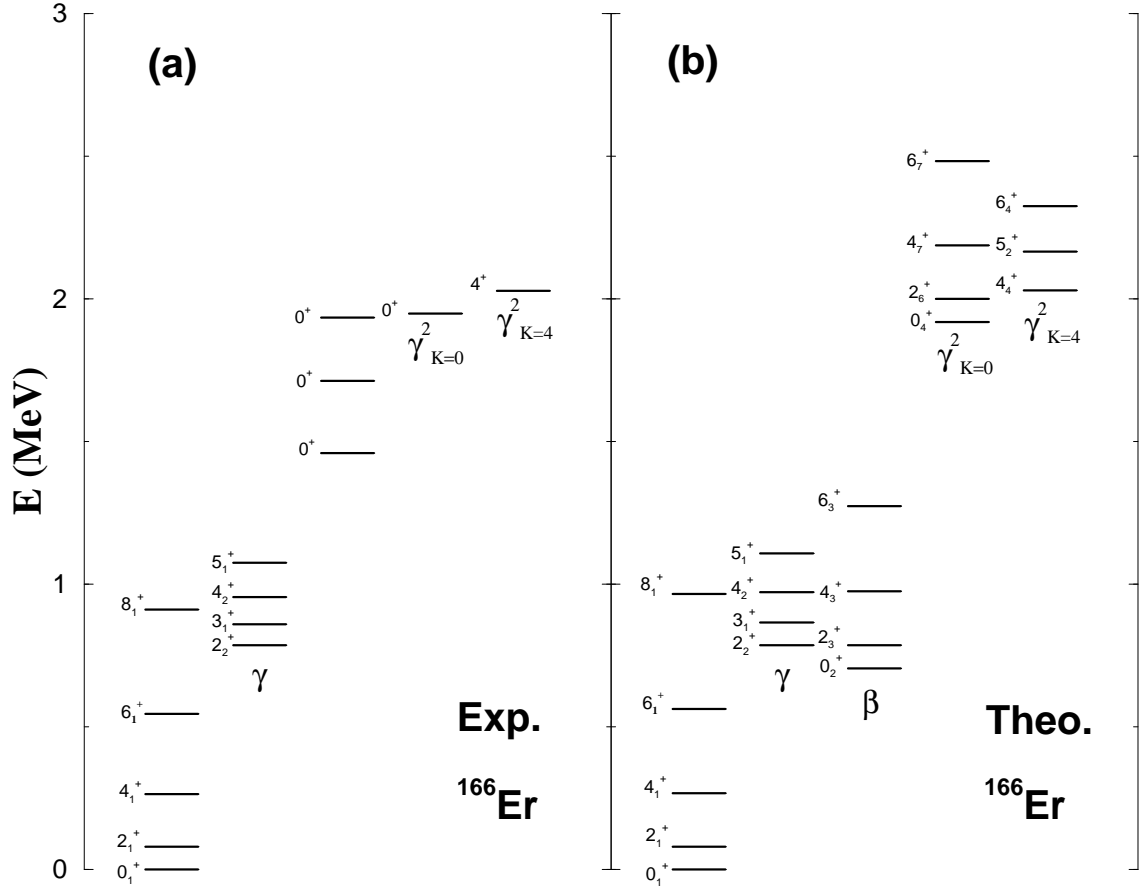,height=15.0cm,angle=-90}}
\end{center}
\caption{Experimental (a) and theoretical (b) spectrum for $^{166}$Er.
The theoretical results are obtained with the Hamiltonian 
$\hat H=
13.43~\hat L^2-
20.84~\hat C_2[SU(3)]+
9.296~10^{-3}~\hat C_2[SU(3)]^2$
(all coefficients in keV)
with $\chi=-\sqrt{7}/2$. 
The boson number is $N=15$.} 
\label{fig-166er-su3}
\end{figure}

Let us consider the case of $^{166}$Er
which is, as already mentioned, one of particular interest
because both double-$\gamma$ excitations
(with $K=0$ and $K=4$) have been identified~\cite{Fahl96,Garr97}. 
To carry out the schematic calculation,
we use the Hamiltonian~(\ref{cas-ham2}) with $b=0$.
The experimental values
for the single and double-$\gamma$ energy ratios
are $R_0^{\gamma}=2.76$ and $R_4^{\gamma}=2.50$
and can be compared directly
with the expressions~(\ref{ratios-c1}-\ref{ratios-c2})
leading two values for $-c/a$,
namely $5\times 10^{-4}$ and $1 \times10^{-4}$.
Both solutions are fairly close
and any value in between them will correctly describe
the anharmonicity of the $K=0$ and $K=4$ bands.
The value of $a$ and the strength of the rotational term
are fixed from the excitation energies
of the $2^+_1$ and $2^+_2$ levels.
After these simple considerations
one arrives at the following Hamiltonian:
\begin{equation}
\label{ham-su3-sch}
\hat H=
13.43~\hat L\cdot\hat L-
20.84~\hat C_2[SU(3)]+
9.296~10^{-3}~\hat C_2[SU(3)]^2, 
\end{equation}
where the coefficients are given in keV.
The theoretical and experimental spectra
are compared in figure~\ref{fig-166er-su3}
and a very good agreement is obtained.
However, due to the simplicity of the calculation,
$\gamma$ and $\beta$ bands are degenerate in energy
which is not the case experimentally. 

To complete the description,
$E2$ transition probabilities must be computed also.
The calculation of $B(E2)$ values
in the $SU(3)$ limit,
as in other situations where degeneracies occur,
must be treated with care 
and an appropriate basis must be chosen
for states with the same energy.
A natural way to do this work is
to slightly lift the degeneracy of the $SU(3)$ Hamiltonian. 
The levels can be split using in the Hamiltonian
a value $\chi=-1.30$ which is very close to its $SU(3)$ value. 
With this $SU(3)$ breaking
the degeneracy is lifted in a natural way
because the $\beta$ band is pushed up in energy,
as is observed.
One may expect that with this small change in $\chi$
the $SU(3)$ spectrum will keep its properties.
The $E2$ transition operator is:
\begin{equation}
\label{t-e2}
\hat T(E2)=
e_{\mbox{eff}}(s^\dag\tilde d+d^\dag\tilde s+
\chi(d^\dag\times\tilde d)^{(2)}).
\end{equation}
The value of $\chi$ that best reproduces the data
is $\chi=-0.26$.
Note that $\chi$
in the $\hat T(E2)$ operator and in the Hamiltonian
is different. 
The effective charge is fixed to reproduce 
$B(E2;2_1^+\rightarrow 0_1^+)$: $e_{\mbox{eff}}^2=(1.97)^2$ W.u..
In table~\ref{table-166er-su3} theoretical and experimental transition rates
involving the ground and the $\gamma$ bands are compared.

This simple analysis suggests
that a four-body operator of the type in~(\ref{ham-su3-sch})
provides a good description of
both single- and double-$\gamma$ bands.
In the next section this schematic analysis
is extended to non-$SU(3)$ situations.

\section{General Hamiltonians with up to four-body interactions}
\label{4b}
A general Hamiltonian with all possible three- and four-body terms
can, in principle, be constructed
but the number of parameters is so high
that a study, even a schematic one,
of the effect of the different terms
on energy spectra and electromagnetic transitions is impossible. 
Schematic IBM Hamiltonians have been used for many years
and, in particular, the quadrupole-quadrupole interaction
has been very successful in describing
a wide variety of nuclear spectra~\cite{Iach87,Cast88}.
On the other hand, in the previous section it was shown
that an expansion in terms of Casimir operators,
which are mainly related to quadrupole operators,
leads to a satisfactory description 
of ground, single- and double-$\gamma$ bands. It is worth noting 
that such an expansion in terms of Casimir operators has been
successfully used in Molecular Physics where spectroscopic data
provide many anharmonic states \cite{Iach95}. 
This leads us to propose a Hamiltonian as a quadrupole expansion
that includes up to four-body terms.
An alternative Hamiltonian can be based on
an expansion in terms of pseudo-Casimir operators,
which we define here as operators
that become a true Casimir operator
only for a particular choice of one structure parameter.
For example, the $\hat C_2[SU(3)]$ operator
is related to the quadrupole operator through  
$2\hat Q\cdot\hat Q+{3\over4}\hat L^2$
where 
$\hat Q=s^\dag\tilde d+d^\dag\tilde s
\pm\sqrt{7}/2(d^\dag\times\tilde d)^{(2)}$.
If the value $\pm\sqrt{7}/2$ is changed to $\chi$,
a new operator $\hat C_2[SU(3)]_\chi$ is obtained
which we refer to as a pseudo-Casimir operator.
It is a Casimir operator only for $\chi=\pm\sqrt{7}/2$. 

Guided by the results of the previous section,
two possible Hamiltonians
that include up to four-body interactions,
can be proposed,
one based on a quadrupole expansion
and the other on a pseudo-Casimir expansion:
\begin{eqnarray}
\label{ham-q} 
\hat H_Q&=&
\kappa'~\hat L\cdot\hat L+
a~\hat Q\cdot\hat Q+
b~(\hat Q\times\hat Q\times\hat Q)^{(0)}+
c~(\hat Q\cdot\hat Q)(\hat Q\cdot\hat Q),\\
\label{ham-pc}
\hat H_{pC}&=&
\kappa'~ \hat L\cdot\hat L+
a~\hat C_2[SU(3)]_\chi+
b~\hat C_3[SU(3)]_\chi+
c~\hat C_2[SU(3)]^2_\chi,
\end{eqnarray}
where $\hat Q=\hat Q(\chi)$ and 
\begin{eqnarray}
\hat C_2[SU(3)]_\chi&=&
2\hat Q(\chi)\cdot\hat Q(\chi)+
{3\over4}\hat L^2,\\
\hat C_3[SU(3)]_\chi&=&
-4\sqrt{35}(\hat Q(\chi)\times\hat Q(\chi)\times\hat Q(\chi))^{(0)}
-{9\over2}\sqrt{15}(\hat L\times\hat L\times\hat Q(\chi))^{(0)}.
\end{eqnarray}
For simplicity and taking into account
the analysis done in the previous section,
in the following $b=0$ is taken
in equations~(\ref{ham-q}) and (\ref{ham-pc}).
   
\subsection{Double-$\gamma$ band heads}
There are three nuclei
that have double-$\gamma$ bands
identified without ambiguity:
$^{164}$Dy, $^{166}$Er, and $^{168}$Er~\cite{Corm97,Fahl96,Garr97,Born91}.
In this section the band heads of these nuclei
are studied using the Hamiltonians~(\ref{ham-q}-\ref{ham-pc})
to get an improved description of the anharmonicity phenomenon. 

The number of bosons for
$^{164}$Dy and $^{168}$Er is $N=16$
while for $^{166}$Er it is $N=15$.
In the different calculations shown in this section 
the parameters of the Hamiltonian have been chosen
as to reproduce as well as possible
not only the heads of single- and double-$\gamma$ bands
but also the structure of the bands. Because these calculations are
schematic and do not try to be the best answer, the parameters of the
Hamiltonian, for simplicity, will not be given fully and only in the
case of final spectra will be shown.
For comparison, the calculation with three-body terms
(see section~\ref{3b} and reference~\cite{Garc00b}) is also included. 

In figures~\ref{fig-164dy-sch},
\ref{fig-166er-sch},
and \ref{fig-168er-sch}
the heads of single- and double-phonon bands are shown.
In each figure six panels are included:
a) experimental data,
b) calculation with Hamiltonian~(\ref{ham-q}) and $\chi=-\sqrt{7}/2$, 
c) calculation with Hamiltonian~(\ref{ham-q}) and $\chi=-0.55$,   
d) calculation with Hamiltonian~(\ref{ham-pc}) and $\chi=-\sqrt{7}/2$,
e) calculation with Hamiltonian~(\ref{ham-pc}) and $\chi=-0.55$, and 
f) calculation with a three-body term
$(d^\dag\times d^\dag\times d^\dag)^{(4)}
\cdot
(\tilde d\times\tilde d\times\tilde d)^{(4)}$.
In each panel, from left to right
are represented the ground state,
the $\gamma$ band head, 
the $\beta$ band head,
the double-$\gamma$ $K=0$ band head,
and the double-$\gamma$ $K=4$ band head. 
Due to the controversy on the nature of the $\beta$ band,
several candidates for the latter have been included. For the same
reason in the figures will be used the label ``$\beta$''. 
In $^{164}$Dy and $^{168}$Er
no information on the double-$\gamma$ $K=0$ band exists.
The value $\chi=-0.55$ is chosen
as an alternative to the $SU(3)$ value
because it describes very well
the $E2$ transition probabilities in this mass region.
Also, the same value of $\chi$ is taken
in the Hamiltonian and in the transition operator,
in line with the Consistent-Q Formalism (CQF)~\cite{Warn82}.
In section~\ref{realistic} this {\it ansatz} is used
in the complete analysis of spectra and $B(E2)$ transitions
for $^{164}$Dy, $^{166}$Er, and $^{168}$Er.

The most striking feature 
of figures~\ref{fig-164dy-sch}--\ref{fig-168er-sch} 
is that in all calculations
the position of the double-$\gamma$ band heads
and the degree of anharmonicity is well reproduced. 
Thus, the energies of the different band heads only
are not sufficient to completely determine the Hamiltonian.
Nevertheless, not all possible terms
are able to create sufficient anharmonicity
in the double-$\gamma$ bands.
For example, in the case of three-body terms,
a phenomenological study of the most relevant type of term
shows that only
$(d^\dag\times d^\dag\times d^\dag)^{(4)}
\cdot 
(\tilde d\times\tilde d\times\tilde d)^{(4)}$
is able to produce the required anharmonicity
(see section~\ref{3b} and reference~\cite{Garc00b}).
On the other hand,
only a few four-body interactions have been explored here
and it cannot be excluded that other four-body terms
can produce the appropriate degree of anharmonicity.      

In order to decide which Hamiltonian is more appropriate,
a description should be attempted not only of band heads
but also of the structure of the bands
and of $E2$ transition probabilities.
 
\subsection{Moments of inertia}
\label{inertia}
\begin{figure}[hbt]
\begin{center}
\mbox{\epsfig{file=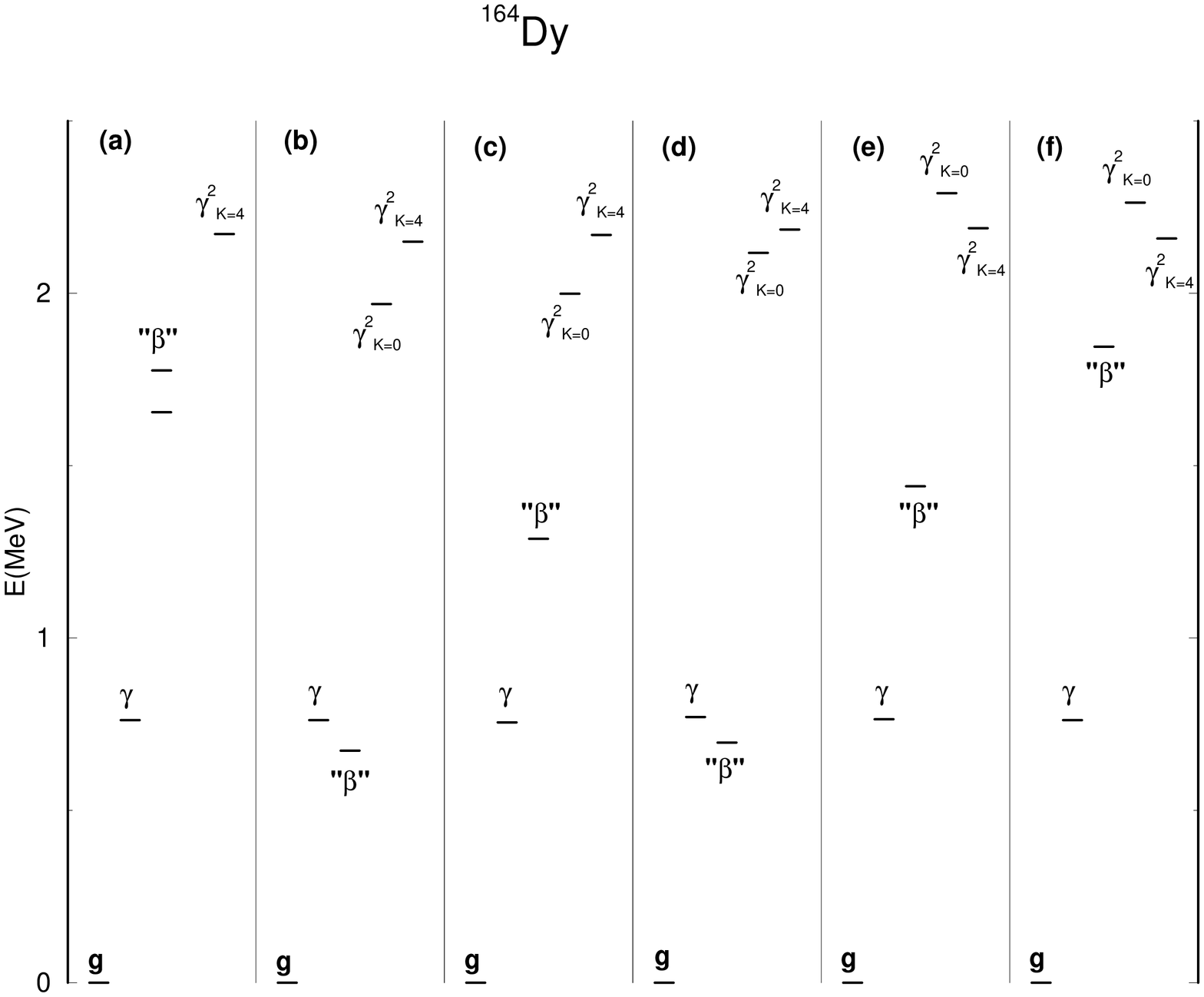,height=15.0cm,angle=-90}}
\end{center}
\caption{Band heads of ground, $\gamma$, $\beta$, double-$\gamma$ $K=0$
and double-$\gamma$ $K=4$ bands of $^{164}$Dy.
Panels correspond to
(a) experimental data,
(b) calculation with Hamiltonian~(\ref{ham-q}) and $\chi=-\sqrt{7}/2$, 
(c) calculation with Hamiltonian~(\ref{ham-q}) and $\chi=-0.55$,   
(d) calculation with Hamiltonian~(\ref{ham-pc}) and $\chi=-\sqrt{7}/2$,
(e) calculation with Hamiltonian~(\ref{ham-pc}) and $\chi=-0.55$, and 
(f) calculation with Hamiltonian~(\ref{ham}).}
\label{fig-164dy-sch}
\end{figure}

\begin{figure}[hbt]
\begin{center}
\mbox{\epsfig{file=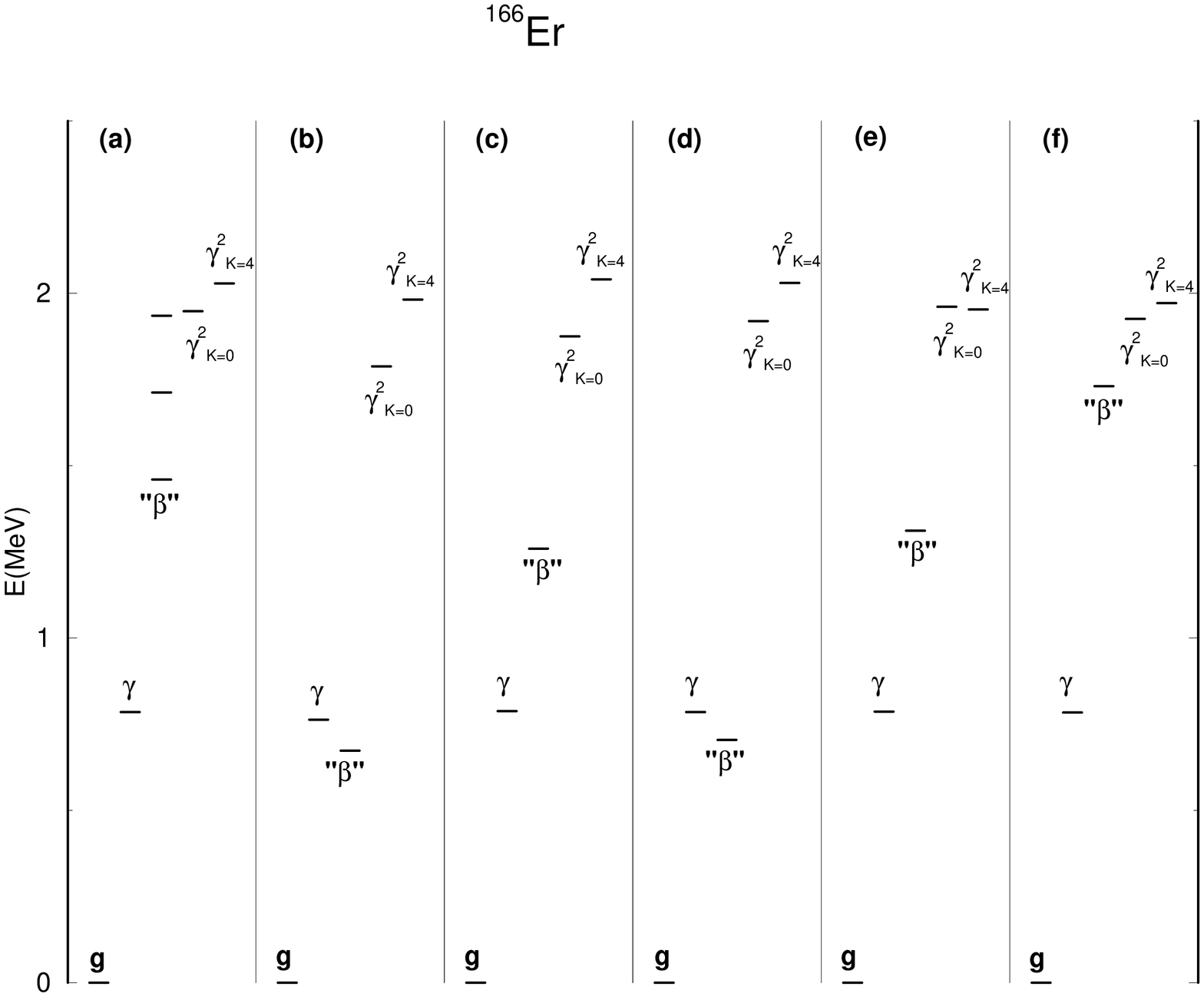,height=15.0cm,angle=-90}}
\end{center}
\caption{Band heads of $^{166}$Er.
See caption of figure~\ref{fig-164dy-sch}.} 
\label{fig-166er-sch}
\end{figure}

\begin{figure}[hbt]
\begin{center}
\mbox{\epsfig{file=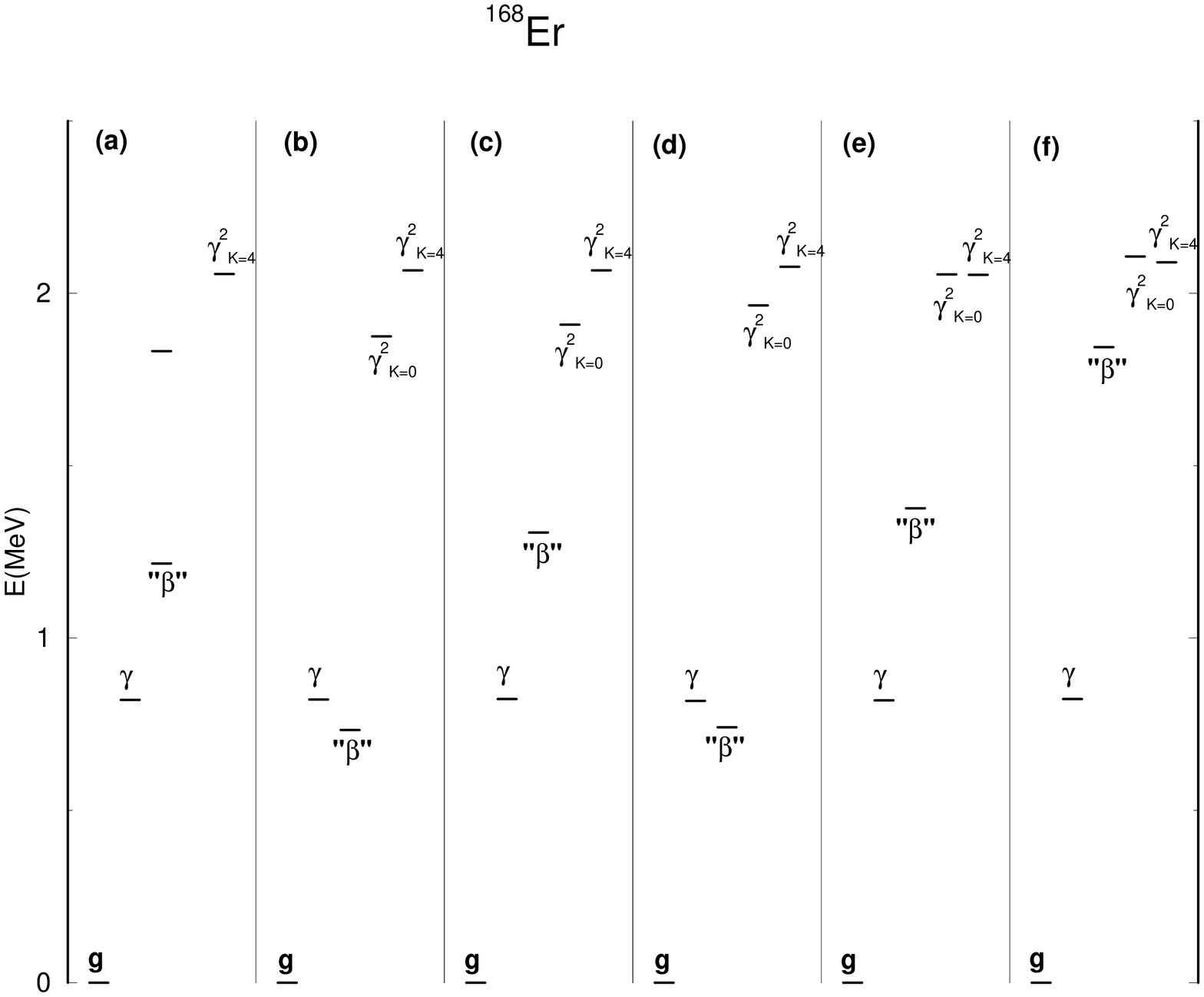,height=15.0cm,angle=-90}}
\end{center}
\caption{Band heads of $^{168}$Er.
See caption of figure~\ref{fig-164dy-sch}.} 
\label{fig-168er-sch}

\end{figure}

The study of the moments of inertia of the lowest bands
is a very sensitive way to test
the different calculations shown
in figures~\ref{fig-164dy-sch}, \ref{fig-166er-sch},
and \ref{fig-168er-sch}.
Particular attention will be paid to the dynamic moment of inertia
which can be obtained from the relation
between angular momentum and $\gamma$-ray energy~\cite{Wu93,Ejir89}
and can be approximated by,
\begin{equation}
\label{m-inertia}
{\cal I}\approx
2\hbar^2{{dJ}\over{dE_\gamma}},
\end{equation}
where $J$ is given dimensionless. In a plot of 
$\gamma$-ray energy versus $J$,
${\cal I}$ will be the slope.
Equation~(\ref{m-inertia}) can be used
to study the structure of the rotational bands
in comparison with experimental results. 

\begin{figure}[hbt]
\begin{center}
\mbox{\epsfig{file=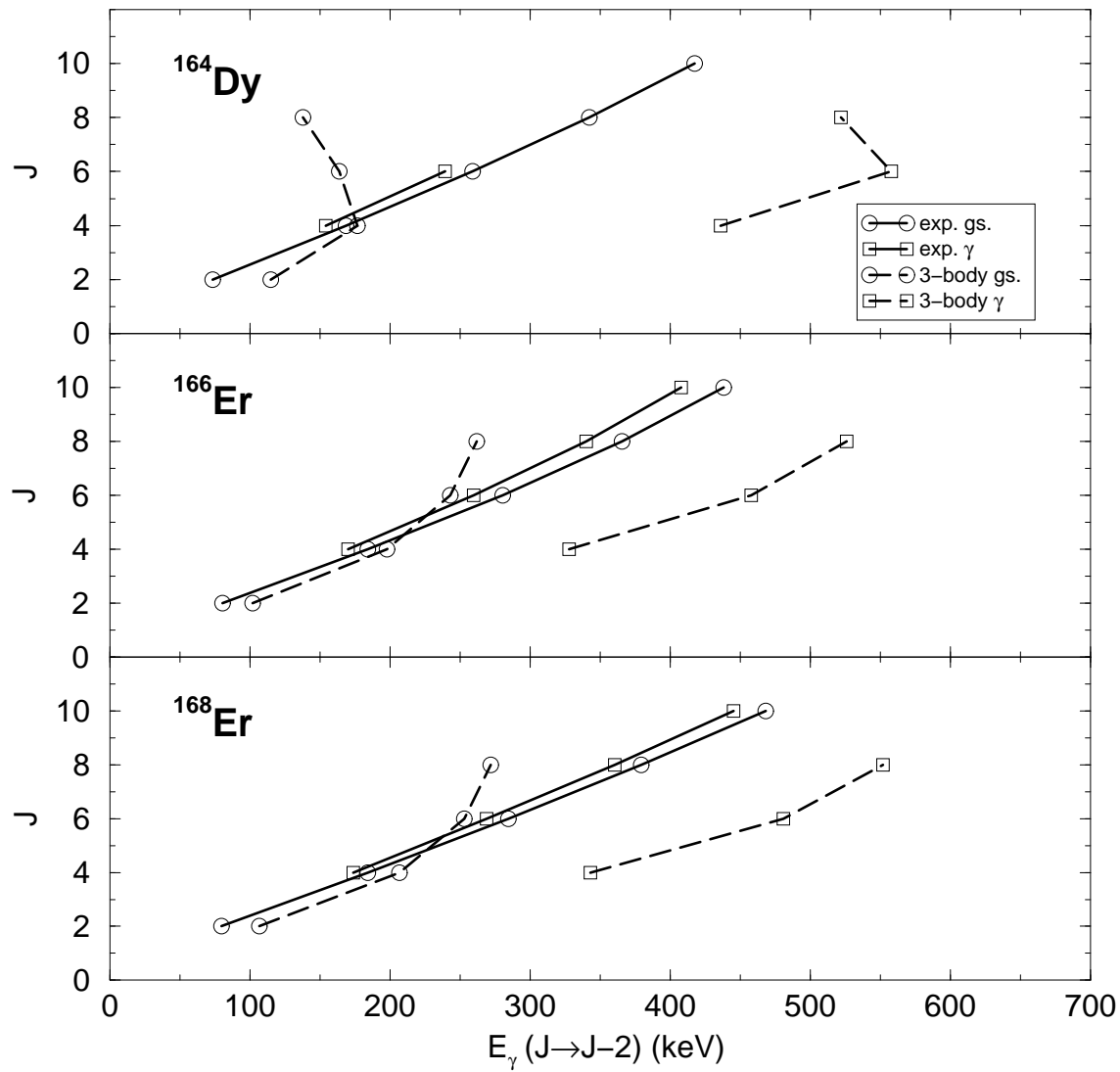,height=15.0cm,angle=-90}}
\end{center}
\caption{Total angular momentum $J$ (dimensionless) of the initial
state versus $\gamma$-ray energies for $\Delta J=2$ transitions
in ground and $\gamma$ bands,
for $^{164}$Dy, $^{166}$Er, and $^{168}$Er.
Full lines correspond to experimental data
and long-dashed lines correspond to the calculated results
obtained with the Hamiltonian~(\ref{ham}). The corresponding
parameters are given in the text.}
\label{fig-iner-3b}
\end{figure}

In figure~\ref{fig-iner-3b}
the moments of inertia of
the ground and $\gamma$-bands
in the nuclei $^{164}$Dy, $^{166}$Er, and $^{168}$Er
are compared to those obtained
with the Hamiltonian of section~\ref{3b}. The parameters for 
$^{164}$Dy are $\kappa=24.2$ keV, $\chi=-0.55$,
$\kappa'=-6.0$ keV, and $\theta_4=51.0$ keV, for $^{166}$Er are
indicated in section \ref{3b} and for $^{168}$Er are 
$\kappa=24.1$ keV, $\chi=-0.55$, $\kappa'=-1.9$ keV, 
and $\theta_4=31.6$ keV.
The predicted moments of inertia disagree completely
with the almost pure rotational structure observed experimentally.
So, although this simple description
based on one single three-body term
describes well the anharmonic position
of the double-phonon band heads,
it fails in the moments of inertia of ground and $\gamma$-bands.
A more realistic description 
needs others three-body terms.
A recent analysis~\cite{unpu} shows, however,
that Hamiltonians with {\em two} different three-body terms
cannot get the correct moment of inertia.   

\begin{figure}[hbt]
\begin{center}
\mbox{\epsfig{file=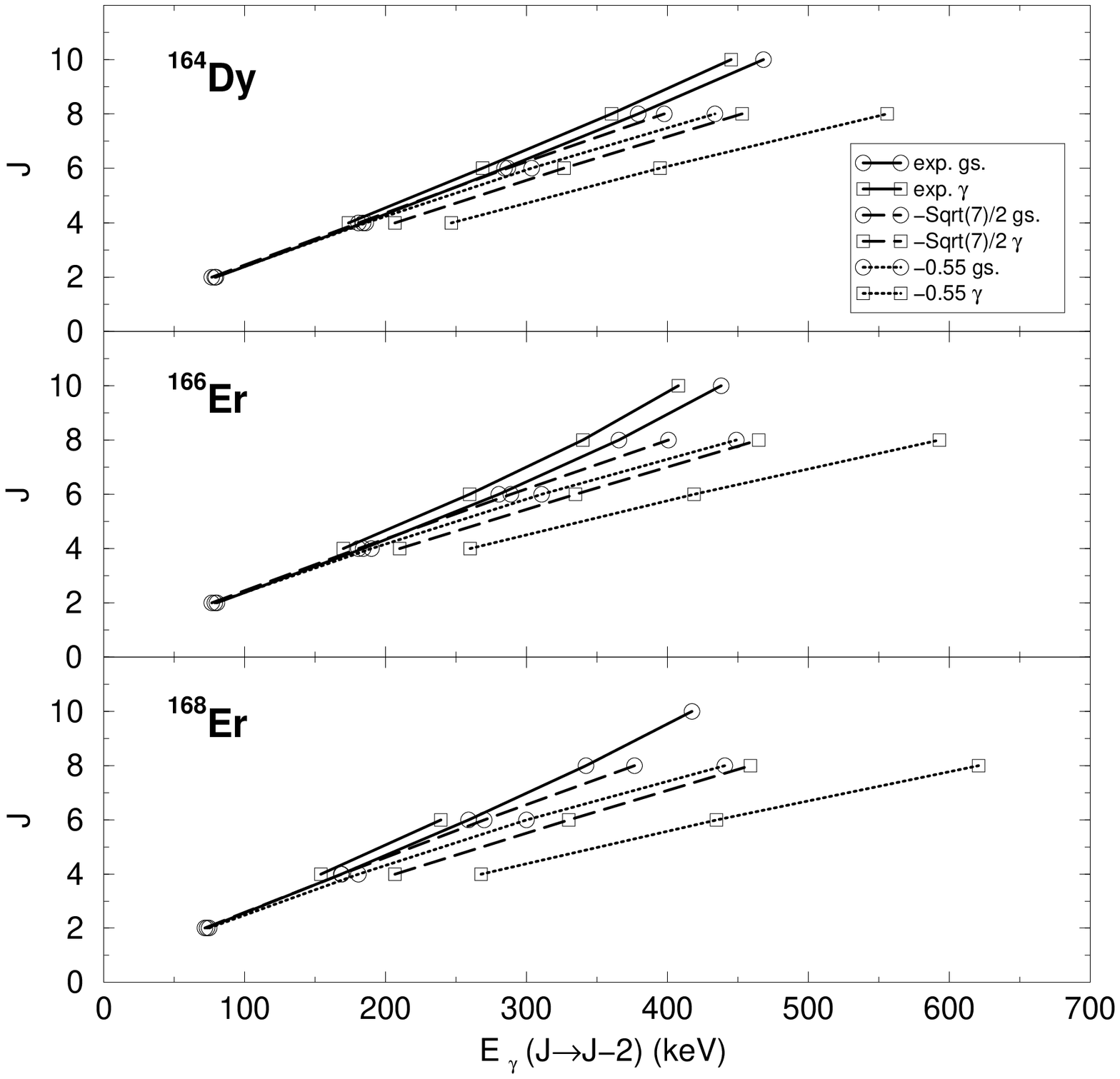,height=15.0cm,angle=-90}}
\end{center}
\caption{Same caption as figure~\ref{fig-iner-3b}
but calculated results obtained with the Hamiltonian~(\ref{ham-q}).}
\label{fig-iner-q}
\end{figure}

\begin{figure}[hbt]
\begin{center}
\mbox{\epsfig{file=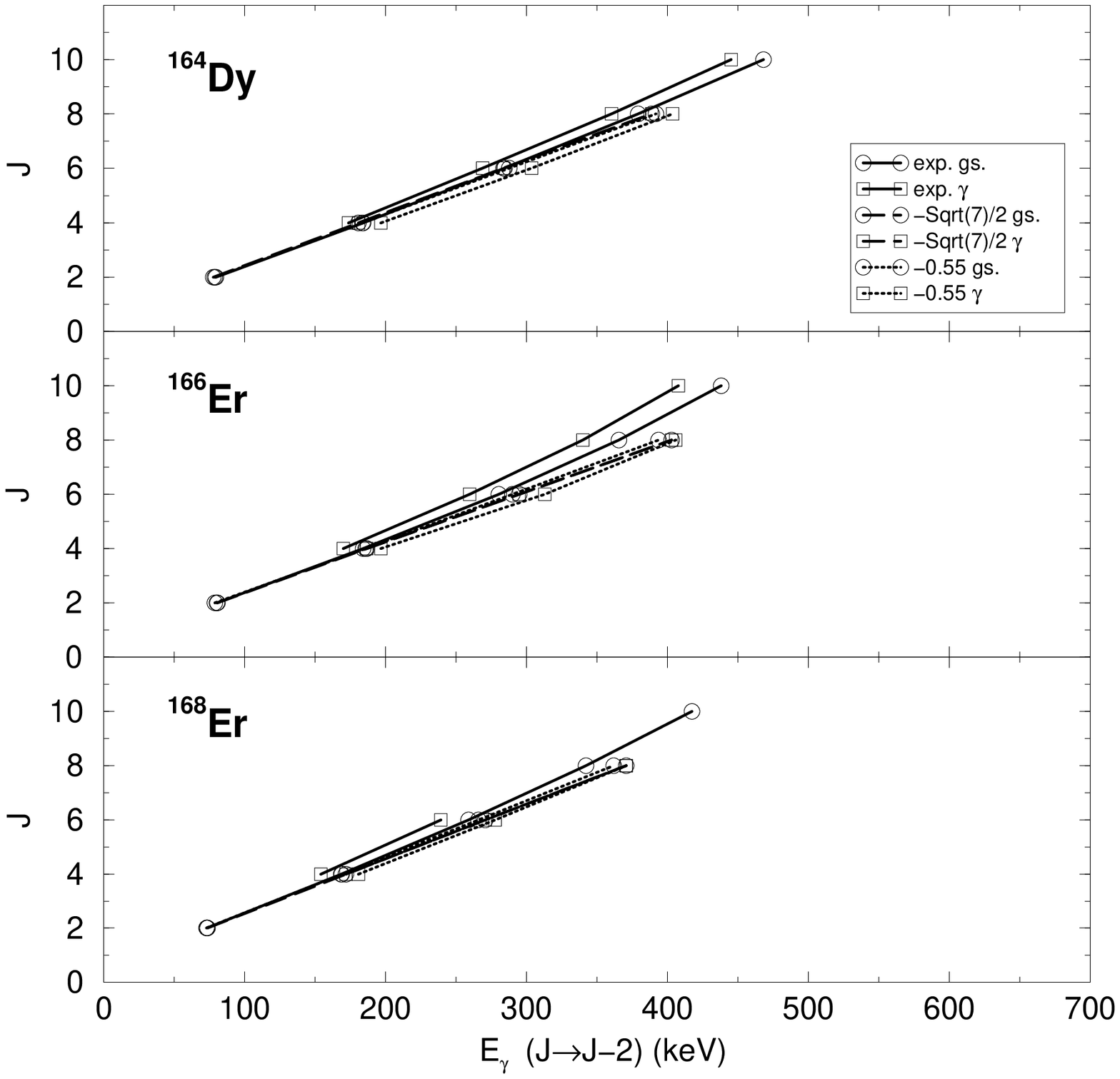,height=15.0cm,angle=-90}}
\end{center}
\caption{Same caption as figure~\ref{fig-iner-3b}
but calculated results obtained with the Hamiltonian~(\ref{ham-pc}).} 
\label{fig-iner-cas}
\end{figure}

The results with the Hamiltonian~(\ref{ham-q})
are shown in figure~\ref{fig-iner-q}. Two different calculations are
shown which correspond to panels $b$ and $c$, respectively, 
of figures \ref{fig-164dy-sch}--\ref{fig-168er-sch}. 
In this case a better agreement is obtained
but still some discrepancies remain,
especially when a realistic value for $\chi$
is taken ($\chi=-0.55$).
Finally, in figure~\ref{fig-iner-cas}
the results with the Hamiltonian~(\ref{ham-pc})
are compared with the data. Again, two different calculations are
shown which correspond to panels $d$ and $e$, respectively, of figures
\ref{fig-164dy-sch}--\ref{fig-168er-sch}.
Here, good agreement is found
both for $\chi=-\sqrt{7}/2$ and $\chi=-0.55$. 

The different results
with the Hamiltonians~(\ref{ham-q}) and (\ref{ham-pc}),
can be understood qualitatively
by analyzing the structure of 
$(\hat Q(\chi)\cdot\hat Q(\chi))(\hat Q(\chi)\cdot\hat Q(\chi))$. 
With just two-body terms
the Hamiltonians~(\ref{ham-q}) and (\ref{ham-pc})
are equivalent. 
This is different when up to four-body terms are included
because $(\hat Q(\chi)\cdot\hat Q(\chi))(\hat Q(\chi)\cdot\hat Q(\chi))$
does not only contribute to $\hat C_2[SU(3)]^2_\chi$ and $\hat L^2$
but also to $(\hat L^2)^2$.
This can be clarified with the equation
\begin{eqnarray}
\label{qqqq}
(\hat Q(\chi)\cdot\hat Q(\chi))(\hat Q(\chi)\cdot\hat Q(\chi))=
{1\over 4}\hat C_2[SU(3)]^2_\chi-{3\over 8}\hat C_2[SU(3)]^2_\chi 
\hat L^2+{9\over 64}(\hat L^2)^2.
\end{eqnarray} 
As a consequence,
$(\hat Q(\chi)\cdot\hat Q(\chi))(\hat Q(\chi)\cdot\hat Q(\chi))$
substantially modifies the rotational structure of a band
even in the case of pure $SU(3)$. 
This is how can be qualitatively understood that
{\em a description in terms of pseudo-Casimir operators
is the most appropriate for dealing with anharmonic vibrations.}
In the next section a complete analysis is given
of the nuclei under study
in the framework of CQF using the Hamiltonian~(\ref{ham-pc}).

\subsection{Realistic calculations}
\label{realistic}
The complete calculated spectra of $^{164}$Dy, 
$^{166}$Er, and $^{168}$Er and the most relevant $E2$ transition 
probabilities are presented in this section. They are compared with
existing data. In each calculation the same value of $\chi$ has been
used both in the Hamiltonian~(\ref{ham-pc}) and in the electromagnetic 
operator~(\ref{t-e2}).
Finally, the effective charge in the transition operator~(\ref{t-e2}), 
$e_{\mbox{eff}}$,
was fixed for each nucleus to reproduce the 
$B(E2;2_1^+\rightarrow 0_1^+)$ value.

\begin{figure}[hbt]
\begin{center}
\mbox{\epsfig{file=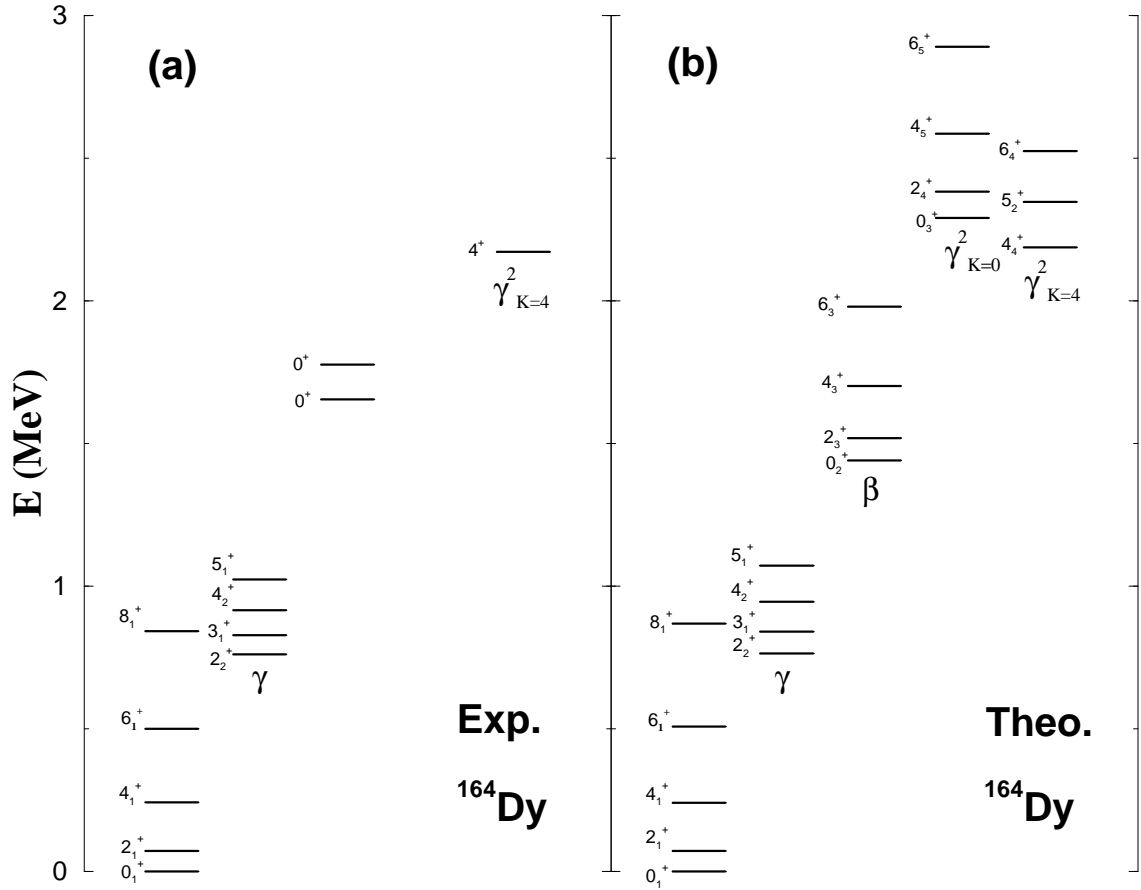,height=15.0cm,angle=-90}}
\end{center}
\caption{Experimental (a) and theoretical (b) spectrum for $^{164}$Dy.
The Hamiltonian~(\ref{ham-pc}) is used
with parameters
$\kappa'=12.18$ keV,
$a=-92.90$ keV,
$c=0.05150$ keV, and
$\chi=-0.55$.
The boson number is $N=16$.} 
\label{fig-164dy-fin}
\end{figure}

\begin{figure}[hbt]
\begin{center}
\mbox{\epsfig{file=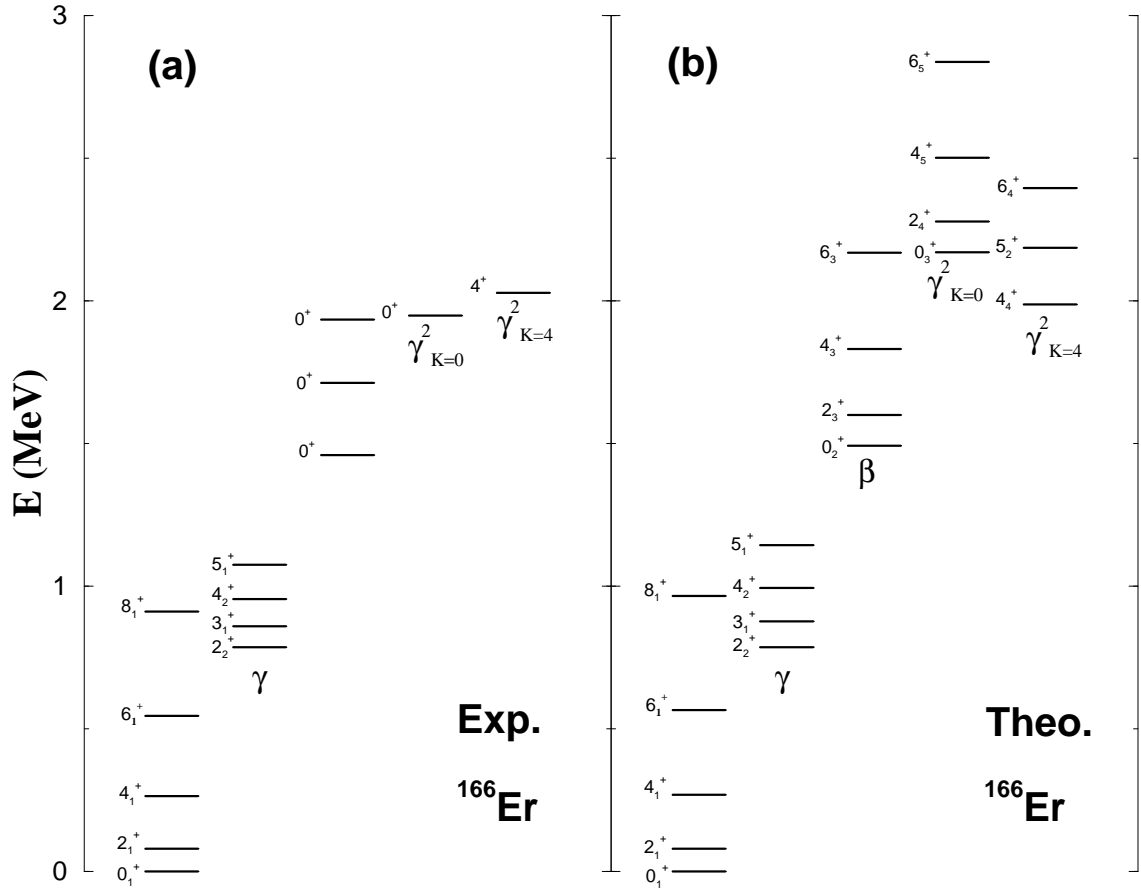,height=15.0cm,angle=-90}}
\end{center}
\caption{Experimental (a) and theoretical (b) spectrum for $^{166}$Er.
The Hamiltonian~(\ref{ham-pc}) is used
with parameters
$\kappa'=13.55$ keV,
$a=-75.40$ keV,
$c=0.05286$ keV, and
$\chi=-0.45$.
The boson number is $N=15$.} 
\label{fig-166er-fin}
\end{figure}

\begin{figure}[hbt]
\begin{center}
\mbox{\epsfig{file=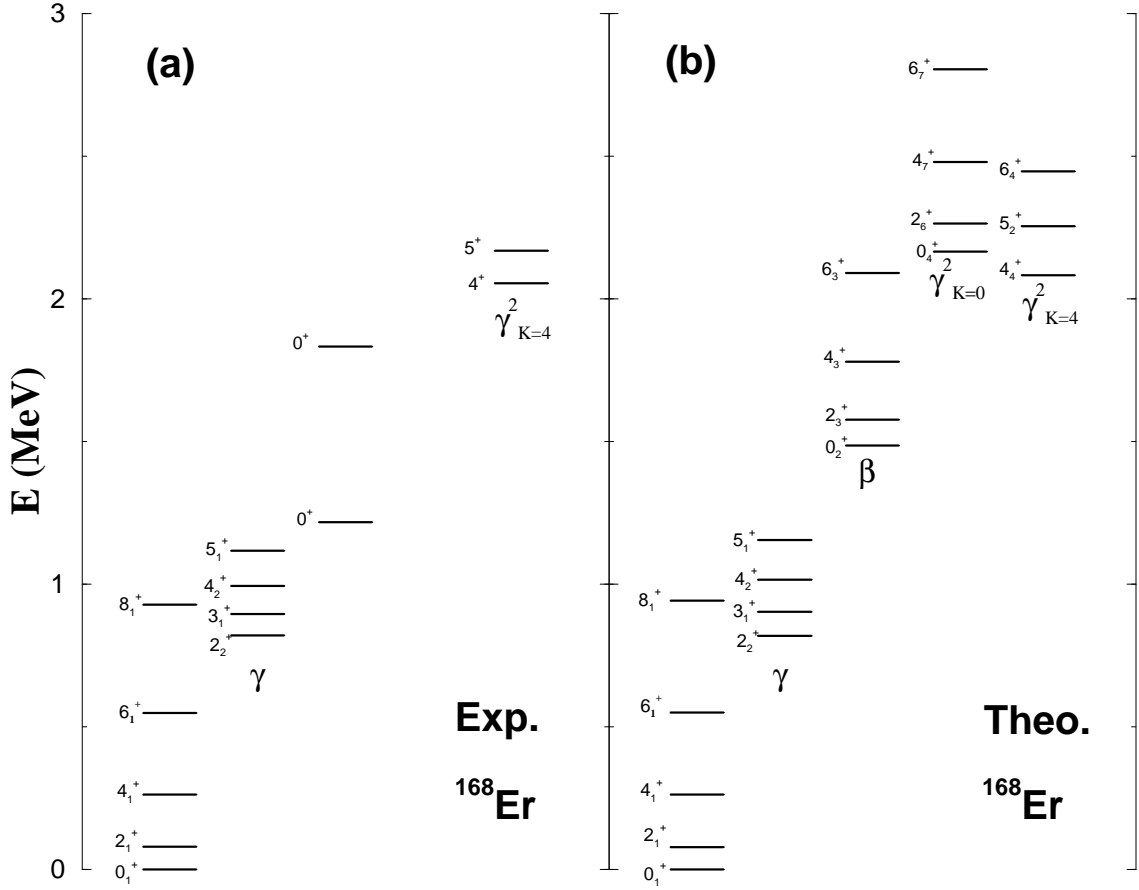,height=15.0cm,angle=-90}}
\end{center}
\caption{Experimental (a) and theoretical (b) spectrum for $^{168}$Er.
The Hamiltonian~(\ref{ham-pc}) is used
with parameters
$\kappa'=13.23$ keV,
$a=-67.25$ keV,
$c=0.04080$ keV, and
$\chi=-0.50$.
The boson number is $N=16$.} 
\label{fig-168er-fin}
\end{figure}

The experimental and calculated spectra
for $^{164}$Dy, $^{166}$Er, and $^{168}$Er are shown in 
figures~\ref{fig-164dy-fin}, \ref{fig-166er-fin}, and
\ref{fig-168er-fin}, respectively. The parameters used in the
calculations are listed  in table~\ref{table-ham}.
The parameters that yield the best fit to the energy spectra
are very similar in the three cases
which is consistent with the analysis carried out
in the preceding sections.
The overall description of the energies is satisfactory.
The calculated low-lying bands
are in good agreement with their experimental counterparts (see also
figure \ref{fig-iner-cas})
while the double-$\gamma$ band-head energies
are close to the experimental values.
The calculated ratios~(\ref{ratios}) are:
$R_0^\gamma=3.31$ and $R_4^\gamma=2.81$ for $^{164}$Dy,
$R_0^\gamma=3.08$ and $R_4^\gamma=2.43$ for $^{166}$Er, and  
$R_0^\gamma=2.93$ and $R_4^\gamma=2.46$ for $^{168}$Er,
to be compared with the experimental ones:
$R_4^\gamma=2.84$ for $^{164}$Dy, 
$R_0^\gamma=2.82$ and $R_4^\gamma=2.45$ for $^{166}$Er, and  
$R_4^\gamma=2.50$ for $^{168}$Er.
Only for the $\gamma^2_{K=0}$ vibration
the experimental and theoretical results are slightly different
in the sense that this framework
overestimates the anharmonic behavior for the $\gamma^2_{K=0}$ band. 
This can be corrected
by increasing the value of $|\chi|$ in the Hamiltonian
which, however, will introduce one more parameter
because in the electromagnetic operator
a different value of $\chi$ must be used. 

For the calculation of $E2$ transition probabilities
the same $\chi$ values as in the Hamiltonian are adopted.
The effective charges are
$e^2_{\mbox{eff}}=(1.66)^2$ W.u., 
$e^2_{\mbox{eff}}=(1.83)^2$ W.u., and 
$e^2_{\mbox{eff}}=(1.67)^2$ W.u.,
for $^{164}$Dy, $^{166}$Er, and $^{168}$Er respectively.
In tables~\ref{table-164dy-fin},
\ref{table-166er-fin},
and \ref{table-168er-fin}
the observed $B(E2)$ values and ratios
concerning $\gamma$-vibrational states
are compared with the theoretical results.
In general, a good overall agreement is obtained
in the three cases under study.

\section{Conclusions}
\label{conclu}
In this paper the problem of anharmonicity
in the $\beta$ and $\gamma$ vibrations of deformed nuclei 
was addressed in the context of the interacting boson model.
The occurrence or not of anharmonicity
was shown to be related to
the order of the interactions between the bosons
and the conclusions of the analysis
can be summarized as follows.
If the Hamiltonian includes up to two-body interactions,
no sizeable anharmonicity can be obtained
and the observed behavior cannot be obtained. 
The origin of this behavior
is related to the existence of a first-order phase transition
between rotational and vibrational nuclei~\cite{Iach87}
which excludes any interplay between one- and two-body terms,
necessary to obtain anharmonic spectra.
For up to three-body interactions that preserve $SU(3)$ symmetry
it can be {\em shown} that
the observed anharmonicity cannot be fully reproduced.
Furthermore, extensive numerical calculations
indicate that even a general IBM Hamiltonian
that includes up to three-body interactions
has difficulty in reproducing all observed aspects
of ground, single- and double-$\gamma$ bands.
However, due to the large parameter space
of three-body interactions
which is difficult to search exhaustively,
this cannot be considered as a firm conclusion
and alternative approaches ({\it e.g.}~mean-field)
should be tried to tackle the same problem.
Finally, it was shown
that a simple parametrization of the IBM Hamiltonian
that includes up to four-body interactions
can account for all observed properties
of the three deformed nuclei
with firmly established double-$\gamma$ vibrations. 
In particular,
the introduction of $SU(3)$ pseudo-Casimir operators
allows to describe the double-phonon states
while keeping correct the properties of low-lying bands.

It should be emphasized that,
in spite of its fourth-order character,
the Hamiltonian considered here
is only slightly more complex than the usual IBM Hamiltonian
(one more parameter)
and is a straightforward extension
of the Consistent-Q Formalism
that was previously successfully applied to many nuclei.
Also clear from our study
is the need for more experimental information
about the double-phonon vibrations in deformed nuclei
beyond the three cases known at present:
our analysis indeed has shown
that this information represents a challenging test
of any theoretical description of deformed nuclei.

\section*{Acknowledgements}
We are grateful to  C.E.~Alonso, K.~Heyde and, F.~Iachello 
for valuable comments. 
Two of the authors (PVI and JEGR) wish to thank the
Institute for Nuclear Theory, University of Washington,
where this work was initiated. One of the authors (JEGR) thanks the
F.W.O.~for financial support. This work was supported in part by the
Spanish DGICYT under project number PB98-1111.

\begin{table}
\caption{Observed and calculated $B$(E2) values and ratios for
$^{166}$Er in a schematic calculation using an $SU(3)$ Hamiltonian.
The E2 operator~(\ref{t-e2}) is used with
$e_{\rm eff}^2=(1.97)^2$ W.u.\
and $\chi=-0.26$.}
\begin{tabular}{lcc}
&\multicolumn{2}{c}{$B$(E2) value or ratio}\\
\cline{2-3}
&Observed&Calculated\\
\hline
$B({\rm E}2;2_1^+\rightarrow0_1^+)$ (W.u.)&$214\pm10$ $^a$ &214\\
$B({\rm E}2;4_1^+\rightarrow2_1^+)$ (W.u.)&$311\pm10$ $^a$ &302\\
$B({\rm E}2;2_\gamma^+\rightarrow0_1^+)$ (W.u.)&$5.5\pm0.4$ $^a$ &5.4\\

${\displaystyle
{B({\rm E}2;0_{\gamma\gamma}^+\rightarrow2_\gamma^+)\over
 B({\rm E}2;2_\gamma^+\rightarrow0_1^+)}}$&$3.8\pm1.3$ $^b$ 
($2.2{{\textstyle +1.1}\atop{\textstyle -0.7}}$ $^c$)&2.5\\

${\displaystyle
{B({\rm E}2;4_{\gamma\gamma}^+\rightarrow2_\gamma^+)\over
 B({\rm E}2;2_\gamma^+\rightarrow0_1^+)}}$&$1.3\pm0.4$ $^b$
($0.9\pm0.3$ $^c$)&2.5\\
\end{tabular}

$^a$ From reference~\cite{Shur92}.

$^b$ From reference~\cite{Garr97}. 

$^c$ From reference~\cite{Fahl96}.

\label{table-166er-su3}
\end{table}

\begin{table}
\caption{Parameters of the Hamiltonian~(\ref{ham-pc})
obtained in the best fit to spectra and $B(E2)$ transitions
in the nuclei $^{164}$Dy, $^{166}$Er, and $^{168}$Er.}
\begin{tabular}{llllll}
Nucleus   &$\kappa'$ (keV)&$a$ (keV)&$c$ (keV)&$\chi$&N \\
\hline
$^{164}$Dy& 12.18         &-82.90   & 0.05150 &-0.55 &16 \\
$^{166}$Er& 13.55         &-75.40   & 0.05286 &-0.45 &15 \\
$^{168}$Er& 13.23         &-67.25   & 0.04080 &-0.50 &16   
\end{tabular}
\label{table-ham}
\end{table}

\begin{table}
\caption{Observed and calculated $B$(E2) values and ratios for $^{164}$Dy.
The E2 operator~(\ref{t-e2}) is used with
$e_{\rm eff}^2=(1.66)^2$ W.u.\
and $\chi=-0.55$.}
\begin{tabular}{lcc}
&\multicolumn{2}{c}{$B$(E2) value or ratio}\\
\cline{2-3}
&Observed&Calculated\\
\hline
$B({\rm E}2;2_1^+\rightarrow0_1^+)$ (W.u.)&$209\pm3$ $^a$ &209\\
$B({\rm E}2;4_1^+\rightarrow2_1^+)$ (W.u.)&$272\pm14$ $^a$ &298\\
$B({\rm E}2;2_\gamma^+\rightarrow0_1^+)$ (W.u.)&$4.0\pm0.4$ $^a$ &3.9\\

${\displaystyle
{B({\rm E}2;4_{\gamma\gamma}^+\rightarrow2_\gamma^+)\over
 B({\rm E}2;2_\gamma^+\rightarrow0_1^+)}}$& 
$0.5-3.9$ $^b$ &3.3\\
\end{tabular}

$^a$ From reference~\cite{Shur92b}.

$^b$ From reference~\cite{Corm97}.
\label{table-164dy-fin}
\end{table}

\begin{table}
\caption{Observed and calculated $B$(E2) values and ratios for $^{166}$Er.
The E2 operator~(\ref{t-e2}) is used with
$e_{\rm eff}^2=(1.83)^2$ W.u.\
and $\chi=-0.45$.}
\begin{tabular}{lcc}
&\multicolumn{2}{c}{$B$(E2) value or ratio}\\
\cline{2-3}
&Observed&Calculated\\
\hline
$B({\rm E}2;2_1^+\rightarrow0_1^+)$ (W.u.)&$214\pm10$ $^a$ &214\\
$B({\rm E}2;4_1^+\rightarrow2_1^+)$ (W.u.)&$311\pm10$ $^a$ &304\\
$B({\rm E}2;2_\gamma^+\rightarrow0_1^+)$ (W.u.)&$5.5\pm0.4$ $^a$ &5.9\\

${\displaystyle
{B({\rm E}2;0_{\gamma\gamma}^+\rightarrow2_\gamma^+)\over
 B({\rm E}2;2_\gamma^+\rightarrow0_1^+)}}$&$3.8\pm1.3$ $^b$ 
($2.2{{\textstyle +1.1}\atop{\textstyle -0.7}}$ $^c$)&1.8\\

${\displaystyle
{B({\rm E}2;4_{\gamma\gamma}^+\rightarrow2_\gamma^+)\over
 B({\rm E}2;2_\gamma^+\rightarrow0_1^+)}}$&$1.3\pm0.4$ $^b$
($0.9\pm0.3$ $^c$)&2.7\\
\end{tabular}

$^a$ From reference~\cite{Shur92}.

$^b$ From reference~\cite{Garr97}

$^c$ From reference~\cite{Fahl96}
\label{table-166er-fin}
\end{table}

\begin{table}
\caption{Observed and calculated $B$(E2) values and ratios for $^{168}$Er.
The E2 operator~(\ref{t-e2}) is used with
$e_{\rm eff}^2=(1.67)^2$ W.u.\
and $\chi=-0.50$.}
\begin{tabular}{lcc}
&\multicolumn{2}{c}{$B$(E2) value or ratio}\\
\cline{2-3}
&Observed&Calculated\\
\hline
$B({\rm E}2;2_1^+\rightarrow0_1^+)$ (W.u.)&$207\pm10$ $^a$ &207\\
$B({\rm E}2;4_1^+\rightarrow2_1^+)$ (W.u.)&$318\pm10$ $^a$ &294\\
$B({\rm E}2;2_\gamma^+\rightarrow0_1^+)$ (W.u.)&$4.80\pm0.17$ $^a$ &4.6\\

${\displaystyle
{B({\rm E}2;4_{\gamma\gamma}^+\rightarrow2_\gamma^+)\over
 B({\rm E}2;2_\gamma^+\rightarrow0_1^+)}}$&$0.5-1.6$ $^b$ 
&2.7\\

${\displaystyle
{B({\rm E}2;5_{\gamma\gamma}^+\rightarrow3_\gamma^+)\over
 B({\rm E}2;2_\gamma^+\rightarrow0_1^+)}}$&$0.7-3.5$ $^b$ 
&2.1\\
\end{tabular}

$^a$ From reference~\cite{Shir94}.

$^b$ From reference~\cite{Born91}.
\label{table-168er-fin}
\end{table}
\end{document}